\documentclass[a4paper,10pt]{article}
\usepackage[english]{babel}\usepackage[cp1250]{inputenc}
\usepackage{amsfonts,amsmath,graphicx}
\usepackage{comment}
\numberwithin{equation}{section}
\usepackage{color}

\newcommand{\e}[1]{{\rm e}^{#1}}

\newcommand{\sgn}{{\rm sgn}}
\newcommand{\s}{{\rm s}}
\def\c{{\rm c}}
\def\d{{\rm d}}

\begin{document}
\usefont{OT1}{pbk}{m}{n}
\newcommand{\hm}{\underline{\bf ??? Ugh ???}}


%
%

\title{
Cosmologically inspired Kastor--Traschen solution
}

\author{{\sc Martin \v{C}erm\'{a}k}$^a$ and {\sc Martin~Zouhar}$^b$
\\~\\
$^{a}$Department of Physical Electronics, \\Faculty of Science, Masaryk University Brno, \\Kotl\'{a}\v{r}sk\'{a} 2, Brno, Czech Republic\\~\\
$^{b}$CEITEC - Central European Institute of Technology,\\ c/o Masaryk University Brno,\\ Kamenice 5, CZ 625 00 Brno, Czech Republic}




\maketitle


\begin{abstract}
Kastor--Traschen (KT) type solution in a cosmological set up is studied in this article.
We examine a hybrid of a KT metric and a Friedmann-Robertson-Walker-Lemaitre (FRWL) solution.
The problem is treated in a general number of dimensions $D~\geq~4$ and we include the cosmological constant $\Lambda$ parameter into the Einstein--Maxwell equations.

The matter source consists of two fluids -- charged dust and neutral fluid with non-vanishing pressure.

The equations of motion for the fluid and electromagnetic (EM) field are written down and an exact solution generalising the extremely charged Reissner--Nordstr\"{o}m black hole to an arbitrary spatial curvature parameter is presented.
We examine metric and singularities of curvature scalars, trapped horizons as well as energy conditions.
\section*{Keywords}
Kastor--Traschen, \and Friedmann-Robertson-Walker-Lemaitre, \and exact solution, \and general relativity, \and spherical symmetry.
\end{abstract}

\section{Introduction}
Since the formulation of general relativity by Albert Einstein in the early twentieth century, scientists have been trying to find exact solutions to the Einstein equations, equations of motion for the metric field (describing the space--time) and matter sources, in various settings.
Some of them include cosmological solutions and solutions describing various matter source distributions that are finite in space.
A natural desire has emerged -- to find a solution describing an object immersed in a cosmological space--time.

There are several solutions, such as the McVittie one~\cite{mcvittie} and metric ans\"{a}tze that may describe such a situation, e.g. the Lemaitre--Tolman--Bondi metrics~\cite{Griffiths2009}.

Dynamic black hole solutions are objects that have been intensively studied in recent years, e.g. generating theorems~\cite{Kothawala:2004fy} and conservation laws~\cite{Hayward:2006ss},~\cite{Hayward:2005ej}.

\medskip
FRWL space-time \cite{Horsky2004} is another exact solution of the Einstein equations, which describes, both globally and spatially, isotropic space-time and represents one of the most famous cosmological solutions.

\medskip
Majumdar--Papapetrou (MP) solutions describe static space--times of a charged fluid with mass and charge densities in equilibrium~\cite{Hartle:1972ya}, i.e. the repulsive electrostatic forces are balanced by attracting gravitational forces.

It turns out that such solutions have a simple form.
The metric is determined by a single function that is directly related to the non--relativistic electrostatic potential of the charge distribution in question.
The MP solutions can be extended to the case of a positive cosmological constant; such solutions are generally called KT solutions~\cite{Brill:1993tm}.

\medskip
In this article, we follow the above mentioned line of research.
We examine a KT type solution in a cosmological FRWL inspired setting by considering co--moving charged dust and perfect fluid as the matter source.

\medskip
Outline of this article is as follows:
We introduce the metric of interest, hybrid KT--FRWL metric, in Section~\ref{behold.the.Metric}.
Einstein--Maxwell equations are examined in Section~\ref{Einstein.and.Max} starting from a general charged fluid source and applying it to the metric considered; the equations are reduced to a set of basic equations on which the rest of this article is based on.
In Section~\ref{no.shells}, we comment on $\Lambda$--electro-vacuum charged dust shells related to a proposition from our earlier article~\cite{Cermak:2012pt}.

The remaining Sections~\ref{basic.pre.analysed}~--~\ref{energy.conditions}, except for the inevitable conclusion in Section~\ref{few.final.words}, are devoted to the examination of various aspects (such as singularities, trapped horizons and energy conditions) of possible solutions with special emphasis on a spherical symmetry.

\section{Kastor-Traschen--FRWL (KT--FRWL) metric}\label{behold.the.Metric}

The notation is as follows:
Lower-case Greek indices run from $0$ to $D-1,$ lower-case Latin indices from $1$ to $D-1.$
$x^0=ct,$ where $t$ is the time coordinate, with $c$ being the speed of light; the Lorentzian signature is mostly minus; comma and semicolon denote partial and covariant derivatives; overdot is used for partial derivative w.r.t. time.

\medskip
Let us briefly mention the FRWL metric that is a sub--case of our starting ansatz~\eqref{MP.FRWL.metric}.
The cosmological FRWL space--time is characterised by line element
\begin{equation}
\d s^2 = (c\,\d t)^2 - F^2
h^2\delta_{ij}\d x^i\d x^j
,\; F = F(t),\;
h^{-1} = 1 + \frac{1}{4}kr^2,\, r^2 = 
\delta_{ij}x^i x^j
.\label{13}\end{equation}
An alternate parametrisation of scale factor $F$ is $F~=~e^{a(t)}.$

FRWL space--time is conformally flat, i.e. it can be written -- in a special coordinate system -- in the form
\begin{equation}
g_{\alpha\beta} = \Omega^2\eta_{\alpha\beta}
,\label{17}\end{equation}
where $\eta_{\alpha\beta}$ is Minkowski metric tensor and $\Omega$ is a general non--vanishing function of coordinates $x^{\gamma}.$
We note a space--time is conformally flat if and only if its Weyl tensor (trace--free part of Riemann tensor) vanishes.

The FRWL metric is characterised by scale factor $F(t),$ see~\eqref{13}, and a constant -- scalar curvature parameter $k.$
The latter can be normalised so that it can take values $k\in\{0,\,\pm1\}.$
$k$ determines curvature of space.
In the case of $k=1$ the space is positively curved (e.g. a sphere or hypersphere), in the case of $k=0$ space is flat and in the case of $k=-1$ space is negatively curved (e.g. hyperboloid).

\bigskip
The hybrid KT--FRWL metric considered is~\footnote{
The explicitly written factor '2' in the exponents at the metric functions $\texttt{G}$ ensures the metric coefficients have the correct signs.
}
\begin{equation}
\d s^2 = \texttt{G}^{-2}(c\,\d t)^2 - \texttt{G}^{2/(D-3)}\,\gamma_{ij}\d x^i\d x^j,\;
\texttt{G} = \texttt{G}(t, x^i)
,\label{MP.FRWL.metric}\end{equation}
where the spatial metric $\gamma_{ij}$ will later be chosen to describe a maximally symmetric space with an added flat sector, i.e.
\begin{equation}
\gamma_{ij}\d x^i\d x^j =
\underbrace{h^2[\d r^2+r^2\d\Omega^2_{\beta}]}_{
\displaystyle\mathbb{S}^{(k)}_{\beta+1}}
+ \underbrace{\delta_{AB}\d z^A\d
z^B}_{\displaystyle\mathbb{R}^{D-2-\beta}}
,\label{metric.k.and.flat}\end{equation}
where the coordinates are labelled as
$$\d x^1\equiv \d r;\; \d x^{\Gamma+1}\equiv\d \theta_{\Gamma}, \;\Gamma \in (1,\ldots,\beta);\; \d x^{A+\beta+1}\equiv\d z^{A}, \;A \in (1,...,D-2-\beta)$$
and $\d\Omega^2_{\beta} = \sum\limits^{\beta-1}_{\Gamma=1}\left[
\prod\limits^{\Gamma}_{\Upsilon=1}\sin^2\left(\theta_{\Upsilon}\right)\right]
\left(\d\theta_{\Gamma+1}\right)^2+\d\theta^2_1\,$~denotes volume element of a unit radius sphere of dimension $\beta.$

Product and sum over empty sets of indices are understood as follows
\begin{equation}
\prod_{\emptyset} = 1,\; \sum_{\emptyset} = 0
.\label{empty.set.operations}\end{equation}

The metric $\gamma_{ij}$ is the direct sum of a metric on $(\beta~+~1)$ dimensional space $\mathbb{S}^{(k)}_{\beta+1}$~\footnote{
By metric on $\mathbb{S}^{(k)}_{\beta+1}$ we mean a metric which we obtain from the flat Euclidean metric on $\mathbb{S}_{\beta}\times\mathbb{R},$ multiplied by $h^2$ in a certain coordinate system.
The metric on $\mathbb{S}^{(1)}_{\beta+1}$ is a positive constant curvature metric on $\mathbb{S}_{\beta+1}$ in the case of $k=1.$
}
of constant curvature $k\in\{0,\pm 1\}$ and a flat metric on $\mathbb{R}^{D-2-\beta}.$
It is a simple generalisation of the flat ansatz considered in~\cite{Cermak:2012pt}.

\medskip
If we set both $\dot{\texttt{G}}~=~0$ and $k~=~0,$ then~\eqref{metric.k.and.flat} reduces to the MP metric.

On the other hand, the conformally flat FRWL metric is recovered in the case of both $\texttt{G}(t, x^i)~\Rightarrow~\texttt{G}(t)$ and $\beta~=~(D-2).$
The following time coordinate transformation
$$\texttt{G}(t)^{\frac{1}{(D-3)}} =
F(\tilde{t}),\; \frac{\d t}{\texttt{G}} = \d\tilde{t}$$
brings the metric into a rather standard FRWL space--time form~\eqref{13}.

Both KT and KT--FRWL metrics are not conformally flat in general.

\section{Einstein--Maxwell equations}\label{Einstein.and.Max}
\subsection{Co--moving charged fluid matter source}
The standard General Relativity with Einstein--Hilbert Lagrangian and geometry described by the Levi--Civita connection (i.e. Riemann tensor and its contractions are determined solely by the metric tensor $g_{\alpha\beta}$ and its derivatives) gives the Einstein equations
\begin{equation}
R_{\alpha\beta}-\frac{1}{2}g_{\alpha\beta}R\equiv
G_{\alpha\beta}=KT_{\alpha\beta}+\Lambda g_{\alpha\beta}
.\label{1}\end{equation}
The appearing symbols have the following meaning;
$K=\frac{D-2}{D-3}\frac{4\pi\kappa}{c^4}$ is coupling constant in a $D$ dimensional
space-time.
$T_{\alpha\beta}$ is the total stress-energy-momentum tensor and $\Lambda$ is the cosmological constant.

The matter source considered consists of two different parts -- a charged fluid and an electro--magnetic (EM) field, i.e.
\begin{equation}
T_{\alpha\beta}=\rho_{\rm m}c^2u_{\alpha}u_{\beta}-\Pi_{\alpha\beta}+\varepsilon_0c^2\left(\frac{1}{4}F^{\gamma\delta}F_{\gamma\delta}g_{\alpha\beta}
-F_{\alpha}\;^{\delta}F_{\beta\delta}\right)
,\label{2}\end{equation}
where $\Pi_{\alpha\beta}~=~\Pi_{\beta\alpha}$ is pressure term that satisfies $\Pi_{\alpha\beta}u^{\beta}~=~0.$

In the case of isotropic--pressure fluid, it is given by a simple formula $\Pi_{\alpha\beta}~=~p[g_{\alpha\beta}~-~u_{\alpha}u_{\beta}],$
where $p$ is the (isotropic) pressure.

Both fluid sources (KT charged dust and FRWL fluid) have been written in a single formula~\eqref{2} as we shall assume that the corresponding velocities are parallel.

\medskip
The EM field is described by gauge 1-form potential $A_{\alpha}$ that enters the equation~\eqref{2} through the (strength) electromagnetic tensor \linebreak $F_{\alpha\beta}=A_{\beta;\alpha}-A_{\alpha;\beta}.$
The equation of motion for the EM field, the Maxwell equation, relates the strength tensor $F_{\alpha\beta}$ to electric current $j^{\alpha}_{\rm e}$ as
\begin{equation}
{F}^{\alpha\beta}_{\;\;\;\; ;\alpha} = \frac{1}{\sqrt{|g|}}\left(
\sqrt{|g|}{F}^{\alpha\beta}\right)_{,\alpha} =
\frac{j_{\rm e}^{\beta}}{\varepsilon_0c^2},\label{3}
\end{equation}
where we have used identity relating covariant divergence of an anti-symmetric tensor with "ordinary" divergence.

\bigskip
Let us assume the fluid is at rest in a certain coordinate system.
As a result, the vector of velocity becomes $u_{\alpha}~=~\sqrt{g_{00}}\delta^0_{\alpha}$ and the electric current has only one non--zero component $j_{\rm e}^0=\frac{c\rho_{\rm e}}{\sqrt{g_{00}}}$ in that frame, where $\rho_{\rm e}$ denotes charge density.

Using the expression for the charge current components, the Maxwell equation~\eqref{3} simplifies to
\begin{equation}
(\sqrt{|g|}{F}^{\alpha0})_{,\alpha}=\frac{\sqrt{|g|}}{\sqrt{g_{00}}}\frac{\rho_{\rm e}}{\varepsilon_0c},\;\;
(\sqrt{|g|}{F}^{\alpha i})_{,\alpha}=0
.\label{4}\end{equation}
A suitable ansatz of the potential $A_{\mu},$ that can satisfy the above set of equations~\eqref{4}, is
\begin{equation}
A_{\mu} =
\left[\frac{\phi_{\rm e}}{c}g_{00} + \dot{f}_{\rm G}(t)\right]\delta^0_{\mu}
,\label{A.ansatz}\end{equation}
where $\phi_{\rm e}$ is the electrostatic potential and the addition due to $f_{\rm G}$ represents a gauge $A_{\mu}\rightarrow~A_{\mu}+f_{{\rm G},\mu}$ degree of freedom.

The ansatz~\eqref{A.ansatz} leads to $F_{0i}=-F_{i0}$ being the only non--zero components of the field strength tensor.
This, together with the fact that the fluid is co--moving, in turn implies the off-diagonal components of the stress-energy-momentum tensor are
\begin{equation}
T_{0i} = T^{\rm EM}_{0i} = 0,\;
\left.T_{ij}\right|_{i\neq j} =
\left.T^{\rm EM}_{ij}\right|_{i\neq j} = - \varepsilon_0c^2A_{0,i}A_{0,j}\texttt{G}^2
.\label{co.moving.Tensor}\end{equation}
The diagonal components of the EM contribution become
$$T^{\rm EM}_{00} = - \frac{1}{2}\varepsilon_0c^2g^{ij}A_{0,i}A_{0,j},\;\;
T^{\rm EM}_{ii} = - \varepsilon_0c^2\texttt{G}^2(A_{0,i})^2
-T^{\rm EM}_{00}g_{ii}g^{00}.$$

\bigskip
The Einstein--Maxwell equations are often supplemented by so called energy conditions~\cite{wald2010general} that are the necessary conditions for a mathematical solution to the equations to be physically reasonable.

%
%

\subsection{The Riemann, Weyl and Einstein tensors}
Firstly, let us define some useful abbreviations,
$$\xi_{D}\equiv\frac{1}{D-3},\; \alpha_{DI}\equiv(D-I)\xi_{D}=\frac{D-I}{D-3},\;
\beta_{DI}\equiv (D-I)(D-I-1).$$

In order to write down the Einstein equations, we need to calculate the Riemann, Ricci and Einstein tensor of sub-metric $\gamma_{ij}$ appearing in~\eqref{MP.FRWL.metric}.
Its components are
\begin{equation*}
\def\arraystretch{1.5}\begin{array}{lcl}
\gamma_{11}&=& h^2,\\
\gamma_{ii}&=&r^2h^2\prod\limits^{i-2}_{\Gamma=1}\sin^2(\theta_{\Gamma}),\; i\in(2,\ldots,\beta+1),\\
\gamma_{ii}&=&1,\; i\in(\beta+2,\ldots,D-1).
\end{array}
\end{equation*}

The independent nonzero component of the Riemann tensor can be written in the form
\begin{equation}
\def\arraystretch{1.5}\begin{array}{lcl}
R_{(\gamma)}\!^{i}_{jij}=k\gamma_{jj},\; i\neq j,\; i\wedge j\in(1,..,\beta+1),\\
R_{(\gamma)}\!^{i}_{jij}=0, \; i\vee j\in(\beta+2,..,D-1),
\end{array}
\label{with.flatland1}\end{equation}
with no sum over repeated index $i.$

The Ricci tensor and Ricci scalar take the form
\begin{equation}
\def\arraystretch{1.5}\begin{array}{lcl}
R_{(\gamma)ii}&=&k\beta \gamma_{ii},\; i\in(1,..,\beta+1),
\\
R_{(\gamma)ii}&=&0, \; i\in(\beta+2,..,D-1),
\\
R_{(\gamma)}&=&k\beta(\beta+1).
\end{array}
\label{with.flatland2}\end{equation}

The Einstein tensor of $\gamma_{ij}$ metric can be written
\begin{equation}
\def\arraystretch{1.5}\begin{array}{lcl}
G_{(\gamma)ij} &=& -\frac{1}{2}\beta(\beta - 1)k\gamma_{ij},\;
i,j\leq \beta+1
,\\
G_{(\gamma)ij} &=& -\frac{1}{2}\beta(\beta + 1)k\gamma_{ij},\;
i,j > \beta+1
.\end{array}
\label{with.flatland}\end{equation}

\bigskip
The non--zero Riemann tensor components that are not related by index symmetries of the total space--time metric~\eqref{13} are
\begin{eqnarray*}
R^0_{\; i0j} &=& \xi_{D}\left[\texttt{G}\ddot{\texttt{G}}+ \xi_{D}\dot{\texttt{G}}^2\right]\texttt{G}^{2\xi_{D}}\gamma_{ij}
\\
&&
+\left[\texttt{G}\nabla_{(\gamma)i}\nabla_{(\gamma)j}\texttt{G} - 2\alpha_{D2}\texttt{G}_{,i}\texttt{G}_{,j}
+ \xi_{D}\left(\nabla_{(\gamma)}\texttt{G}\right)^2\gamma_{ij}\right]\texttt{G}^{-2}
,\\
R^i_{\; 0jk} &=& 2\xi_{D} \texttt{G}^{-1}\delta^i_{[k}\dot{\texttt{G}}_{,j]}
,\\
R^i_{\; jkm}
&=& R^{\quad i}_{(\gamma)\; jkm} + 2\xi_{D}\left[\dot{\texttt{G}}^2\texttt{G}^{2\xi_{D}}
- \left(\nabla_{(\gamma)}\texttt{G}\right)^2\texttt{G}^{-2}\right]\delta^i_{[k}\gamma_{m]j}
\\
&&
+ 4\xi_{D}\gamma^{ip}\gamma_{[j|[m}\left(
- \texttt{G}\nabla_{(\gamma)k]}\nabla_{(\gamma)|p]}\texttt{G}
+ \alpha_{D2}\texttt{G}_{,k]}\texttt{G}_{,|p]}\right)\texttt{G}^{-2}
.\end{eqnarray*}
Similarly, the Weyl tensor components
\begin{equation}
W_{\mu\nu\rho\tau} = R_{\mu\nu\rho\tau}
+ \frac{4}{D-2}g_{[\mu|[\tau}R_{\rho]|\nu]}
- \frac{2}{(D-1)(D-2)}Rg_{\mu[\tau}g_{\rho]\nu}
\label{Weyl.defined}\end{equation}
of the space--time described by metric~\eqref{13} are
\begin{eqnarray*}
W_{0i0j} &=&
- \frac{1}{D-2}\left[
R_{(\gamma)ij}-\frac{1}{D-1}R_{(\gamma)}\gamma_{ij}\right]\texttt{G}^{-2}
+ A_{ij}\texttt{G}^{-2}
,\\
W_{ijkm} &=&
- W_{(\gamma)ijkm}\texttt{G}^{2\xi_{D}} - 4\xi_{D}\gamma_{[i|[m}A_{k]|j]}\texttt{G}^{2\xi_{D}}
,\end{eqnarray*}
with abbreviations
$$A_{ij} = B_{ij} - \frac{1}{D-1}\left(\gamma^{mn}B_{mn}\right)\gamma_{ij},\;
B_{ij} = \frac{\texttt{G}\nabla_{(\gamma)i}\nabla_{(\gamma)j}\texttt{G} - (2D-5)\xi_{D} \texttt{G}_{,i}\texttt{G}_{,j}}{\texttt{G}^2}$$
and $W_{(\gamma)ijkm}$ is given by~\eqref{Weyl.defined} with the Riemann tensor (and corresponding contractions) ${R}_{\mu\nu\rho\tau}$ replaced by ${R}_{(\gamma)ijkl},$ the dimensional factors remaining the same
\begin{eqnarray*}
W_{(\gamma)ijij} &=& k\frac{
(D-\beta-1)(D-\beta-2)
}{\beta_{D1}}\gamma_{ii}\gamma_{jj},\; i\wedge j\leq\beta+1
,\\
W_{(\gamma)ijij} &=& k\frac{\beta(\beta+2-D)}{\beta_{D1}}\gamma_{ii},\; i\leq\beta+1<j
,\\
W_{(\gamma)ijij} &=& k\frac{\beta(\beta+1)}{\beta_{D1}},\; \beta+1<i\wedge j.
\end{eqnarray*}

We have defined indices (anti)symmetrisation notation
$$T_{\mu(\nu\rho)} = \frac{1}{2}\left(T_{\mu\nu\rho} + T_{\mu\rho\nu}\right),\;
T_{\mu[\nu|\rho|\tau]} = \frac{1}{2}\left(
T_{\mu\nu\rho\tau} - T_{\mu\tau\rho\nu}\right),$$
i.e. square brackets denote antisymmetrisation, as opposed to round bracket symmetrisation, and indices enclosed within $|\ldots|$ are not affected by the index-symmetry operations (marked outside of the range $|\ldots|$).

Non--zero components of the Einstein tensor $G_{\mu\nu}$ corresponding to the metric~\eqref{13} are
\begin{eqnarray*}
G_{00} &=&
- \frac{1}{2}\alpha_{D2}\left[2\texttt{G}\left(\Delta_{(\gamma)}\texttt{G}\right) - \left(\nabla_{(\gamma)}\texttt{G}\right)^2\right]\texttt{G}^{-4-2\xi_{D}}
\\
&&
+ \frac{1}{2}R_{(\gamma)}\texttt{G}^{-2-2\xi_{D}}
+ \frac{1}{2}\alpha_{D1}\alpha_{D2}\dot{\texttt{G}}^2\texttt{G}^{-2}
,\\
G_{0i} &=& -\alpha_{D2}\texttt{G}_{,0i}\texttt{G}^{-1}
,\\
G_{ij} &=& - \frac{1}{2}\alpha_{D2}\left[2\texttt{G}_{,i}\texttt{G}_{,j} - \left(\nabla_{(\gamma)}\texttt{G}\right)^2\gamma_{ij}\right]\texttt{G}^{-2}
\\
&&
+ G_{(\gamma)ij}
-\frac{1}{2}\alpha_{D2}\left[2\ddot{\texttt{G}}\texttt{G} + \alpha_{D1}\dot{\texttt{G}}^2\right]\texttt{G}^{2\xi_{D}}\gamma_{ij}
,\end{eqnarray*}
where we have introduced $\gamma_{ij}$ related quantities, square of a gradient of a function and Laplace operator,
$$\left(\nabla_{(\gamma)}\texttt{G}\right)^2 = \gamma^{mn}\texttt{G}_{,m}\texttt{G}_{,n},\;
\Delta_{(\gamma)}\texttt{G} = \frac{1}{\sqrt{|\gamma|}}\left(
\sqrt{|\gamma|}\gamma^{mn}\texttt{G}_{,n}\right)_{,m},\;
\gamma = \det\left(\gamma_{ij}\right),$$
Einstein tensor $G_{(\gamma)ij}$ and Ricci scalar $R_{(\gamma)}.$

The above formulae hold for arbitrary time independent $\gamma_{ij}$ and they can be viewed as an extension of~\cite{Lemos:2005md} to the case of $\texttt{G}$ depending also on time.

In the case of $\beta=D-2,$ i.e. no flat sector in~\eqref{metric.k.and.flat}, we have to put
\begin{equation}
R_{(\gamma)} = \beta_{D1} k,\; G_{(\gamma)ij} = -\frac{1}{2}\beta_{D2} k\gamma_{ij}
\label{no.flatland}\end{equation}
into the general formulae for the Einstein tensor presented at the beginning of this sub--section.
Notice that the $\beta=D-2$ metric~\eqref{metric.k.and.flat} reduces to the globally isotropic metric which implies that the corresponding Einstein tensor $G_{(\gamma)ij}$ is also isotropic.

\bigskip
The constant curvature sector metric can be written in an alternate form
\begin{equation}
h^2[\d r^2 + r^2\d\Omega^2_{\beta}] =
\d\vartheta^2 + s^2_k(\vartheta)\d\Omega^2_{\beta}
,\label{metric.k.cases}\end{equation}
we have defined the function $\s_k(\vartheta)$ by
\begin{equation}
\s_k(\vartheta)\equiv \frac{1}{\sqrt{k}}\sin\left(\sqrt{k}\,\vartheta\right) =
\left\{\begin{array}{l}
\sinh\vartheta,\; k = -1,\\
\vartheta,\; k = 0,\\
\sin\vartheta,\; k = +1.\\
\end{array}\right.
\label{abbreviate.k}\end{equation}
We can also define the function $\c_k(\vartheta)$
\begin{equation}
\c_k(\vartheta)\equiv \frac{\d\s_k(\vartheta)}{\d\vartheta} =
\s_k(\vartheta)_{,\vartheta}=
\left\{\begin{array}{l}
\cosh\vartheta,\; k = -1,\\
1,\; k = 0,\\
\cos\vartheta,\; k = +1.\\
\end{array}\right.
\label{abbreviate.k1}\end{equation}
The $\s_k$ and $\c_k$ satisfy
\begin{equation}
\c_k^2 + k\s_k^2 = 1,\; \c_{k\, ,\vartheta} = - k\s_k
.\label{extras.k}\end{equation}


\subsection{The basic equations for the KT--FRWL metric}
Let us assume the fluid is at rest in the same coordinate system in which the metric~\eqref{MP.FRWL.metric} is written.

A convenient form of the pressure term $\Pi_{\alpha\beta}$ for the co--moving fluid corresponding to the metric $\gamma_{ij}$~\eqref{metric.k.and.flat} is
\begin{equation}
\Pi_{0\mu} = 0,\;\;
i,j\leq\beta + 1: \Pi_{ij} = p_{\rm k-sector}g_{ij},\;\;
i,j\geq\beta + 2: \Pi_{ij} = p_{\rm flat}g_{ij}
.\label{pressure.term.adapted}\end{equation}

Adopting the ansatz~\eqref{A.ansatz}, it follows that~\eqref{co.moving.Tensor} holds and this allows us to separate time and spatial variables in $\texttt{G}$ and solve $\phi_{\rm e}$ in terms of $\texttt{G}.$

Indeed, the time-space mixed components of the Einstein equations imply
$$0 = G_{0i}\propto \texttt{G}_{,0i}\Rightarrow \texttt{G} = \texttt{T}(t) + \texttt{R}(x^i),$$
i.e. the metric~\eqref{MP.FRWL.metric} simplifies to
\begin{equation}
\d s^2 = \left(\frac{c\d t}{\texttt{G}}\right)^2 -
\texttt{G}^{2\xi_{D}}\gamma_{ij}\d x^i\d x^j,\;
\texttt{G} = \texttt{T}(t)+\texttt{R}(\vec{r})
,\label{14}\end{equation}
where we continue to write $\texttt{G}$ instead of $\texttt{T}~+~\texttt{R}$ for the sake of brevity.

MP metric corresponds to a special case of
$\texttt{T}(t)$ being a constant (that can be scaled so that $\texttt{T}=1$) and $k=0.$

\medskip
The $i~\neq~j$ Einstein equation, see~\eqref{co.moving.Tensor} and the expression for $G_{ij},$ the latter simplified by diagonality of $\gamma_{ij}$~\eqref{metric.k.and.flat} in the coordinate system used, gives
$$\frac{1}{\texttt{G}} = \pm\sqrt{\frac{K\varepsilon_0}{\alpha_{D2}}}\left[
\frac{\phi_{\rm e}}{\texttt{G}^2}
+ c\dot{f}_{\rm G}(t)\right] + \tilde{f}(t),$$
where $\tilde{f}(t)$ is an integration (with respect to spatial coordinates) constant.
We choose such a gauge that the $\dot{f}_{\rm G}(t)$ in the above equation cancels $\tilde{f}(t).$

Then it follows
\begin{equation}
\texttt{G} = \pm\sqrt{\frac{K\varepsilon_0}{\alpha_{D2}}}\phi_{\rm e}
.\label{G.potential}\end{equation}

Using the above results for $\texttt{G}$ and $\phi_{\rm e}(\texttt{G}),$ the remaining Einstein--Maxwell equations determine dependence of $\rho_{\rm m},$ $\rho_{\rm e}$ and $p$ on the metric function $\texttt{G}$
\begin{eqnarray}
\Lambda- Kp_{\rm k-sector} &=&
\alpha_{D2}\ddot{\texttt{G}}\texttt{G} + \frac{\alpha_{D2}\alpha_{D1}}{2}\dot{\texttt{G}}^2
+ \frac{1}{2}\beta(\beta-1)k\texttt{G}^{-2\xi_{D}}
\label{p.curved.Einstein.reduced},\\
p_{\rm flat} &=& p_{\rm k-sector} + \beta kK^{-1}
\label{p.flat.Einstein.reduced},\\
\Lambda+Kc^2\rho_{\rm m} &=&
\frac{\alpha_{D2}\alpha_{D1}}{2}\dot{\texttt{G}}^2
\pm\sqrt{\alpha_{D2}\frac{K}{\varepsilon_0}}\rho_{\rm e}
+ \frac{1}{2}\beta(\beta+1)k\texttt{G}^{-2\xi_{D}}
\label{rho.Einstein.reduced}
\end{eqnarray}
where we have used a non--trivial Maxwell equation
\begin{equation}
\Delta_{(\gamma)}\texttt{G} =
\pm\sqrt{\frac{1}{\alpha_{D2}}\frac{K}{\varepsilon_0}}\rho_{\rm e}\texttt{G}^{\alpha_{D1}}
.\label{Max.reduced}\end{equation}
in the~\eqref{rho.Einstein.reduced}.

Let us note that the Maxwell equation $F^{\mu i}_{\quad ;\mu}~=~0$ is identically satisfied.

The mathematical solution relating $\texttt{G}$ and $\phi_{\rm e}$ or $\rho_{\rm e}$ contains an undetermined sign parameter.
This parameter comes from the fact that corresponding $i~\neq~j$ Einstein-Maxwell equations are quadratic in both gauge-field $A_{\mu}$ and metric function $\texttt{G}$.

In the case of $\beta~=~(D~-~2),$ no flat sector present in $\gamma_{ij},$ the pressure is isotropic $p~=~p_{\rm k-sector}.$

The equations~\eqref{p.curved.Einstein.reduced} and~\eqref{rho.Einstein.reduced} are analogous to the FRWL space--time Einstein equations and for $\phi_{e}=0$, $\texttt{G}=\texttt{G}(t)$ are identical with FRWL Einstein equations.
The equation~\eqref{rho.Einstein.reduced} reduces to the static Majumdar--Papapetrou mass--charge balance condition
in the appropriate limit of all $\Lambda,$ $\dot{\texttt{G}}$ and $k$ vanishing.
The balance condition ensures that attractive gravitational forces are exactly compensated by repulsive electrostatic forces so that the static Majumdar--Papapetrou solution can be obtained.
The metric function $\texttt{G}$ is also a linear function of the electrostatic potential, which itself is a solution to the Poisson equation \eqref{Max.reduced}.

\subsection{Petrov classification of a class of solutions}
We shall use the coordinate system as in~\eqref{metric.k.cases}.
The Petrov classification in higher dimensions~\cite{Milson:2004jx,Pravda:2005qp} is applied.

It relies on finding a Weyl Aligned Null Direction (WAND) which satisfies the same condition as the principal null direction of $D~=~4$ classification, i.e.
\begin{equation}
0 = N_{[\rho\tau][\omega\varphi]}\equiv
k^{\mu}k^{\nu}k_{[\rho}W_{\tau]\mu\nu[\omega}k_{\varphi]},\;
0 = g_{\mu\nu}k^{\mu}k^{\nu}
.\label{WAND.definition}\end{equation}
Then a so called alignment type of the Weyl tensor (with respect to a null basis built using the WANDs)~\cite{Milson:2004jx,Pravda:2005qp} is determined in order to complete the classification.

Let us examine two simple sub--cases satisfying all
\begin{equation}
\texttt{G} = \texttt{G}(t,x^1\equiv\vartheta),\; \left(\Delta_{(\gamma)}\texttt{G}\right)\equiv
\texttt{G}_{,11} + \beta\,\frac{\c_k}{\s_k}\texttt{G}_{,1} = 0,\;
W_{(\gamma)ijkm} = 0
,\label{G.simplify}\end{equation}
one of the two sub--cases represents a solution examined in detail in this article.
Indeed, the third equation in~\eqref{G.simplify} is satisfied e.g. in the two following cases - first, $\beta~=~(D-2),$ $k$ arbitrary and second, $\beta$ arbitrary but $k~=~0$ this time.
The latter case corresponds to the shell solutions examined in~\cite{Cermak:2012pt}.

The first equation in~\eqref{G.simplify} is a natural requirement that the function $\texttt{G}$ respects symmetries of the metric $\gamma_{ij}.$
The first two conditions in~\eqref{G.simplify} imply that $A_{ij},$ introduced in the previous section, takes the form
$$A_{ij} = 
-(\beta + 1)\frac{\c_k}{\s_k}\frac{\texttt{G}_{,1}}{\texttt{G}}\left[\delta^1_i\delta^1_j
-\frac{\gamma_{ij}}{\beta + 1}\right]_{i,j\leq\beta+1}
-(2D-5)\xi_{D}\frac{\left(\texttt{G}_{,1}\right)^2}{\texttt{G}^2}\left[\delta^1_i\delta^1_j
-\frac{\gamma_{ij}}{D-1}\right],$$
where the fact that $\gamma_{11}~=~1$ has been used.
If $i$ or $j$ are greater than $\beta~+~1,$ the first square bracket term in $A_{ij}$ should be omitted.

Thus there are three classes of indices depending on the form of $A_{ij}.$
For that reason, we divide the spatial metric~\eqref{metric.k.and.flat} into three auxiliary sectors - the first one corresponds to $x^1$ only, the second corresponds to the rest of $\mathbb{S}^{(k)}_{\beta+1}$ part (indices labelled by capital Greek letters
) and the last one corresponds to the flat part $\mathbb{R}^{D-2-\beta}$ (indices labelled by capital Latin letters).

Using that~\eqref{G.simplify}, together with the form of $\gamma_{ij}$ in~\eqref{metric.k.and.flat}, implies that $A_{ij}$ is diagonal.
We can define three corresponding auxiliary quantities
\begin{eqnarray*}
\Gamma_i &=& A_{ii}\gamma^{ii},\;\;\; A_{il}k^{l}=A_{il}\gamma^{lm}\hat{k}_{m}=\Gamma_i\hat{k}_{i}, \\
i=1:\, \Gamma_1 &=& -\left[
\beta\frac{\c_k}{\s_k}\frac{\texttt{G}}{\texttt{G}_{,1}} + \xi_{D}(D-2)\frac{2D-5}{D-1}
\right]\frac{\left(\texttt{G}_{,1}\right)^2}{\texttt{G}^2}
,\\
i\in\{2,\ldots,\beta+1\}: \Gamma_i &=& \left[
\frac{\c_k}{\s_k}\frac{\texttt{G}}{\texttt{G}_{,1}} + \xi_{D}\frac{2D-5}{D-1}
\right]\frac{\left(\texttt{G}_{,1}\right)^2}{\texttt{G}^2}
,\\
i\in\{\beta+2,\ldots,D-1\}: \Gamma_i &=& 
\xi_{D}\frac{2D-5}{D-1}\frac{\left(\texttt{G}_{,1}\right)^2}{\texttt{G}^2}
,\end{eqnarray*}
with no sum over the repeated index $i.$
If we further introduce the following quantities
$$\hat{k}_i=\gamma_{ij}k^j,\,
\left|\vec{k}\right|^2=\gamma_{ij}k^ik^j = \hat{k}_jk^j,\,
\Xi_{\rm K} = \frac{A_{ij}k^ik^j}{\left|\vec{k}\right|^2},\,
D_{ij} = (D-2)A_{ij} + \Xi_{\rm K}\gamma_{ij},$$
then the WAND equation~\eqref{WAND.definition} is turned into
\begin{eqnarray}
0 = \frac{4N_{[0i][0j]}}{-\texttt{G}^{4\xi_{D}-2}\left|\vec{k}\right|^2} &=&
\left[\alpha_{D2}(\Gamma_i+\Gamma_j)-\Xi_{\rm K}\right]\hat{k}_i\hat{k}_j
- \xi_{D}\left|\vec{k}\right|^2D_{ij}
\label{double.0},\\
0 = \frac{4N_{[0p][qr]}}{-\texttt{G}^{4\xi_{D}}k_0} &=&
\alpha_{D2}\hat{k}_p\hat{k}_q\hat{k}_r(\Gamma_q-\Gamma_r)
- 2\xi_{D}\left|\vec{k}\right|^2D_{p[q}\hat{k}_{r]}
\label{single.0},\\
0 = \frac{N_{[pq][rs]}}{-\texttt{G}^{6\xi_{D}}\left|\vec{k}\right|^2} &=&
\xi_{D}\hat{k}_{[p}D_{q][r}\hat{k}_{s]}
\label{no.0}
.\end{eqnarray}

We assume that both $\texttt{G}_{,1}$ (in order to exclude "trivial" FRWL metric) and all $\vec{k}$ components appearing in the above equations are non--zero.
This leads to contradictions in many cases thus eliminating unsuitable WAND candidates.

In the following four items, no sum over the repeated indices of the $\vec{k}$'s components is performed; all explicitly appearing components are assumed to be non--zero.
\begin{itemize}
\item
We can start by investigating~\eqref{double.0} in the case of $i~\neq~j$ (so that $D_{ij}~=~0$).
Let us consider e.g. $ij$ given by $\Upsilon_1\Upsilon_2,$\, $\Upsilon_1~\neq~\Upsilon_2,$ and $A_1A_2,$\, $A_1~\neq~A_2,$ i.e. a vector $\vec{k}$ of the form
$$\vec{k} = \underbrace{k^1\partial_1
+ k^{\Upsilon_1}\partial_{\Upsilon_1} + k^{\Upsilon_2}\partial_{\Upsilon_2} + \ldots}_{\displaystyle\mathbb{S}^{(k)}_{\beta+1}\, {\rm part}}
+ \underbrace{k^{A_1}\partial_{A_1} + k^{A_2}\partial_{A_2} + \ldots}_{\displaystyle\mathbb{R}^{D-2-\beta}\, {\rm part}}\;.$$
The dots in the above equation indicate that some other components may be, but need not be, non--zero.

Subtracting the equations corresponding to the choice of indices $\Upsilon_1\Upsilon_2$ and $A_1A_2$ with $\hat{k}$s omitted, reveals
$$0 = 2\alpha_{D2}(\Gamma_{\Upsilon_1}-\Gamma_{A_1})\Rightarrow \texttt{G}_{,1} = 0.$$
Thus a WAND candidate cannot simultaneously have two or more non--zero components with both second and third class indices, irrespective of the value of $k^1.$

\item
WAND candidates with $\vec{k}~=~k^{A_1}\partial_{A_1}~+~k^{A_2}\partial_{A_2}~+~\ldots$ have $\Xi_{\rm K}~=~\Gamma_{A_1}$ and can be ruled out in a similar spirit.

\item
The case of $\beta~=~(D-2)$ is particularly easy to examine because $A_{ij}$ has a unified single form.

Examining the WAND condition $N_{[01][01]}~=~0$ reveals the only admissible WAND candidate has $\vec{k}~=~k^1\partial_1.$
Expressing $\hat{k}^2_1/\left|\vec{k}\right|^2\in(0,1]$ is useful in reaching that conclusion.

This candidate also satisfies the remaining $N_{[\rho\tau][\omega\varphi]}~=~0$ equations, thus it is indeed a WAND.

\item
The WAND candidate list can be shortened to
$$k^1\neq 0, \vec{k}\in\left\{k^1\partial_1,\,
k^1\partial_1 + k^{\Upsilon}\partial_{\Upsilon},\,
k^1\partial_1 + k^A\partial_A\right\}$$
by means of the above listed hints.

%
%
\end{itemize}

The WANDs with $\vec{k}~=~k^1\partial_1$ ($\beta~=~k~=~0$ and $\beta~=~D-2$) are, up to an overall multiplicative factor, null vectors of the form
\begin{equation}
k_{\pm} = \frac{1}{\sqrt{2}}\left[\texttt{G}\;\partial_0
\pm \texttt{G}^{-\xi_{D}}\partial_{1}\right]
.\label{WAND.maximal.beta}\end{equation}
We have chosen the vectors so that $g_{\mu\nu}k^{\mu}_{+}k^{\nu}_{-}~=~1.$

It is straightforward to find out that the alignment type is $(2, 2),$ hence the metric considered and subject to condition~\eqref{G.simplify} is of Petrov type D in the two cases of $\beta~=~k~=~0$ and $\beta~=~(D-2).$

\section{A note on $\Lambda$--electro-vacuum charged dust shells}\label{no.shells}
We have considered $\Lambda$--electro-vacuum charged dust thin shells of KT type in~\cite{Cermak:2012pt} where we have suggested examining the generalisation of the metric ansatz of~\cite{Cermak:2012pt} to the type with $\gamma_{ij}$ as given in~\eqref{metric.k.and.flat}.

\medskip
When relating the proposal to the present study, we are reminded that the charged dust shells find themselves in an otherwise $\Lambda-$electro-vacuum space--time, which means both pressures $p_{\rm flat}$ and $p_{\rm k-sector}$ vanish.

Differentiating the $p_{\rm k-sector}~=~0$ equation~\eqref{p.curved.Einstein.reduced} with respect to $x^i$ and using that $\texttt{G}~=~\texttt{T}(t)~+~\texttt{R}(x^j)$ reveals
$$0 = \texttt{R}_{,i}\left[\alpha_{D2}\ddot{\texttt{T}}
- \frac{\beta(\beta-1)k}{D-3}\left(\texttt{T}+\texttt{R}\right)^{-\alpha_{D1}}\right].$$
If $\texttt{R}_{,i}~=~0,$ we obtain a FRWL metric which we are not interested in.

Consider the nontrivial solution $\texttt{R}_{,i}~\neq~0,$ then the above equation must be satisfied by $[\ldots]~=~0.$
The first term is time dependent only.
The second depends on both time and a spatial coordinate.
In order to ensure $[\ldots]~=~0$ for all admissible values of $x^i,$ we have to impose $\beta(\beta-1)k = 0.$
When this condition is satisfied, the second term in $[\ldots]$ vanishes and so must the first one, $\ddot{\texttt{T}}.$
It implies $\texttt{T}$ is (at most) a linear function of time.
The coefficient at $t$ is linearly related to the FRWL Hubble parameter and we shall simply call it the Hubble constant.

The condition $\beta(\beta-1)k~=~0$ can be split into three different cases.

\begin{itemize}
\item
$k~=~0$

Spatial metric tensor~\eqref{metric.k.and.flat} is Euclidean metric, $\gamma_{ij}=\delta_{ij}.$

\item
$\beta~=~0$

The spatial metric tensor can be transformed into a Euclidean form again, see equation~\eqref{metric.k.cases},
$$\gamma_{ij}\d x^i\d x^j = h^2\d r^2 + \delta_{AB}\d x^A\d x^B =
\d\vartheta^2 + \delta_{AB}\d x^A\d x^B =
\delta_{ij}\d x^i\d x^j.$$

\item
$\beta~=~1$

Assuming both $\beta=1$ and $p=0$ reduces the equation~\eqref{p.flat.Einstein.reduced} to $\frac{k}{K}=0$ which means the only admissible $\gamma_{ij}$ is the Euclidean metric, once again.
\end{itemize}

The above considerations show that the dust shell analysis of~\cite{Cermak:2012pt} cannot be extended to the case of the metric ansatz considered here with non--trivial spatial metric $\gamma_{ij}$~\eqref{metric.k.and.flat}.

In the case of both Euclidean $\gamma_{ij}$ and $p~=~0,$ the basic equations~\eqref{p.curved.Einstein.reduced}, \eqref{p.flat.Einstein.reduced}, \eqref{rho.Einstein.reduced} reduce to
\begin{equation}
\Lambda_{\rm cosm} =
\frac{\alpha_{D2}\alpha_{D1}}{2}\dot{\texttt{G}}^2,\;\;
p_{\rm flat}= p_{\rm k-sector} = 0,\;\;
Kc^2\rho_{\rm m} = \pm
\sqrt{\alpha_{D2}\frac{K}{\varepsilon_0}}\rho_{\rm e}
.\label{shell.basic.reduction}\end{equation}

The first equation in~\eqref{shell.basic.reduction} relates cosmological constant $\Lambda$ and the Hubble constant.
The last equation is the Majumdar-Papapetrou balance condition for charged matter.

\section{Preliminary analysis of the basic equations}\label{basic.pre.analysed}
From now on, we shall consider that the parameter $\beta$ has a maximal value ($\beta=D-2$), i.e. $\gamma_{ij}$ is a maximally symmetric metric~\eqref{metric.k.cases}.

The Einstein--Maxwell equations of co--moving charged fluid in MP--FRWL space--time with the EM field ansatz~\eqref{A.ansatz} have been reduced to a few basic equations that can be written as
\begin{eqnarray}
\texttt{T}(t) + \texttt{R}(\vec{r}) = \texttt{G} &=&
\pm\sqrt{\frac{K\varepsilon_0}{\alpha_{D2}}}\phi_{\rm e},\;
\Delta_{(\gamma)}\texttt{G} =
\pm\sqrt{\frac{1}{\alpha_{D2}}\frac{K}{\varepsilon_0}}\rho_{\rm e}\texttt{G}^{\alpha_{D1}}
\label{Max.deduced}
,\\
\Lambda- Kp &=&
\alpha_{D2}\ddot{\texttt{G}}\texttt{G} + \frac{\alpha_{D2}\alpha_{D1}}{2}\dot{\texttt{G}}^2
+ \frac{1}{2}k\beta_{D2}\texttt{G}^{-2\xi_{D}}
\label{p.Einstein.deduced}
,\\
\Lambda+Kc^2\rho_{\rm m} &=&
\frac{\alpha_{D2}\alpha_{D1}}{2}\dot{\texttt{G}}^2
\pm\sqrt{\alpha_{D2}\frac{K}{\varepsilon_0}}\rho_{\rm e}
+ \frac{1}{2}k\beta_{D1}\texttt{G}^{-2\xi_{D}}
\label{rho.Einstein.deduced}
\end{eqnarray}
in a suitable gauge.

The Maxwell equation~\eqref{Max.deduced} already allows us to comment on the time--dependence of $\rho_{\rm e}.$
Obviously, the $\Delta_{(\gamma)}\texttt{G}$ is time independent and so must be the other side of the equation containing the Laplacian of $\texttt{G}.$
Hence $\pm\rho_{\rm e}~=~f(\vec{r})\texttt{G}^{-\alpha_{D1}}$ for a yet undetermined function $f(\vec{r})~\propto~\Delta_{(\gamma)}\texttt{G}.$
It is evident, that if $\texttt{G}$ is a function of time, then $\rho_{\rm e}$ must be also (unless $f(\vec{r})~=~0$).

\bigskip
Let us split the mass density $\rho_{\rm m}$ into two parts $\rho_{\rm mp}$ (cosmological fluid) and $\rho_{\rm me}$ (KT dust) defined as
\begin{equation}
\rho_{\rm mp}=\rho_{\rm m} - \rho_{\rm me},\;
Kc^2\rho_{\rm me} = \pm\sqrt{\alpha_{D2}\frac{K}{\varepsilon_0}}\rho_{\rm e}
.\label{splitting.balanced}\end{equation}
$\rho_{\rm me}$ is defined via the Majumdar--Papapetrou balance condition.

After splitting, the equation~\eqref{rho.Einstein.deduced} becomes
$$\Lambda+Kc^2\rho_{\rm mp} =
\frac{\alpha_{D2}\alpha_{D1}}{2}\dot{\texttt{G}}^2 + \frac{1}{2}k\beta_{D2}\texttt{G}^{-2\xi_{D}}.$$

\medskip

Let us consider a specific example of $\texttt{T}~=~Hc(t-t_0),$ $H$ being constant, then $\dot{\texttt{T}}\equiv\dot{\texttt{G}}=H$ and $\ddot{\texttt{G}}=0.$
The above equations for $p$ and $\rho_{\rm mp},$ expressing both parameters in terms of $\texttt{G}$ and its derivatives, simplify to
$$\Lambda - Kp =
\frac{\alpha_{D1}\alpha_{D2}}{2}H^2+\frac{k\beta_{D2}}{2\texttt{G}^{2\xi_{D}}},\;
\Lambda+Kc^2\rho_{\rm mp} =
\frac{\alpha_{D1}\alpha_{D2}}{2}H^2 + \frac{k\beta_{D1}}{2\texttt{G}^{2\xi_{D}}},$$
where $H$ is related to the cosmological Hubble parameter $\mathcal{H}$ via $Hc=(D-3)\mathcal{H}$ as follows from definitions of $\mathcal{H}$ and coordinate transformations bringing the metric considered into the standard FRWL form.

A particular solution to the above equations for $p,$ $\rho_{\rm mp}$ and $H$ has the form
\begin{equation}
H = \pm\sqrt{\frac{2[\Lambda + K\lambda]}{\alpha_{D1}\alpha_{D2}}},\;
- [p + \lambda] = \frac{1}{\alpha_{D1}}[c^2\rho_{\rm mp} - \lambda] =
\frac{k\beta_{D2}}{2K\texttt{G}^{2\xi_{D}}}
,\label{Kastor.Traschen.like}\end{equation}
where $\lambda\in\mathbb{R}$ is a constant parameter.
It can be absorbed, from both $p$ and $\rho_{\rm mp},$ into a cosmological constant by a redefinition of $\Lambda~+~K\lambda~\rightarrow~\Lambda$ and corresponding shifts in both $p$ and $\rho_{\rm mp}$ and we shall henceforth set $\lambda~=~0.$

Let us note that the solution in~\eqref{Kastor.Traschen.like} reduces to a Kastor--Traschen solution~\cite{Kastor:1992nn,Brill:1993tm} in the limit $k~=~0.$

\section{A spherically symmetric solution - analysis of the function $\texttt{R}$ and the Misner-Sharp mass}
\label{Extreme.k}
With $\beta~=~(D-2),$ it is natural to assume spherical symmetry of the solution for the metric function $\texttt{G},$ i.e. $\texttt{G}~=~\texttt{T}(t)~+~\texttt{R}(\vartheta)$ in the second coordinate representation of spatial metric~\eqref{metric.k.cases}.

The auxiliary functions $\s_k$ and $\c_k$ defined in~\eqref{abbreviate.k} allow us to write the Poisson--like Maxwell equation as
\begin{equation}
\left(\texttt{R}_{,\vartheta}\left|\s^{D-2}_k(\vartheta)\right|\right)_{,\vartheta} =
\mp
\left|\s^{D-2}_k(\vartheta)\right|\sqrt{\frac{K}{\delta_D\varepsilon_0}}\rho_{\rm e}\texttt{G}^{\alpha_{D1}}
\label{Poisson.k.MP}\end{equation}
which is a natural generalisation of the Poisson equation for KT space-time~\eqref{14}.

The solution of~\eqref{Poisson.k.MP} in the source--free region (characterised by $\rho_{\rm e}=0$) is
\begin{equation}
\texttt{R} = - C_1\int\frac{\d\vartheta}{\left|s^{D-2}_k(\vartheta)\right|}
,\label{18z}\end{equation}
where $C_1$ is a constant of integration.
The integral on the right side of~\eqref{18z} produces another constant of integration.
An analytical solution of the integral in the right hand side of~\eqref{18z} will be derived in the subsequent text.

As in the case of $-\int|x|^{-n}\d x,$ $n~\geq~2$ the primitive function can be found on disconnected domains $(-\infty,0)$ and $(0,\infty)$ and a general formula applicable to both domains -- separately -- is $\sgn(x)/[(n-1)|x|^{n-1}]$ apart from the integration constant.

A primitive function cannot be found on the whole $\mathbb{R}$ because of the integrand being divergent at $x~=~0.$
Nevertheless, we shall write down the primitive functions (related to equation~\eqref{18z}) formulae applicable to both domains because it will be of use later.
This time, the domains will be determined by signum of $\s_k(\vartheta).$

\medskip
Let us introduce abbreviations $I_n~=~\s^{-n}_k,$ $\texttt{R}_n~\equiv~\int~\left|I_n\right|~\d\vartheta$ and denote derivative with respect to the argument of $\s_k$ by prime.
The relation between newly introduced function $\texttt{R}_n$ and function $\texttt{G}$ can be written in the form
$$\texttt{G} = \texttt{T} + \texttt{R} = \texttt{T} - C_1\texttt{R}_{D-2}.$$
Constant of integration originating from $\texttt{R}_{D-2}$ was included in the function $\texttt{T}.$

\medskip
The absolute value in the integrand in the definition of $\texttt{R}_n$ is irrelevant in the case of even $n$ and can be ignored.
If $n$ is odd, the integral is more complicated.

The functions $\s_k(\vartheta)$ are always positive in the intervals
\begin{equation*}
k\in\{0,-1\}:\; \vartheta\in\left[0,\infty\right);\;\;
k = +1:\; \vartheta\in\left[0,\pi\right]
.\label{DF}
\end{equation*}

Irrespective of the $\s_k(\vartheta)$ domain of integration being positive or negative, we can write $\int~\left|I_n\right|~\d\vartheta$ in the form
\begin{eqnarray*}
\texttt{R}_{n-{\rm odd}} \!&=&\!\sgn(\s_k\left(\vartheta\right))\left(
\prod^{\frac{n-1}{2}}_{i=1}\frac{n-2i}{n-2i+1}\ln\left|\frac{\s_k\left(\frac{\vartheta}{2}\right)}{\c_k\left(\frac{\vartheta}{2}\right)}\right|
k^{\frac{n-1}{2}}\right.
\\
&-&\left. \sum ^{\frac{n-1}{2}}_{J=1}
\left[\prod^{\frac{n-1}{2}-J}_{i=1}\frac{n-2i}{n-2i+1}\right]\frac{k^{\frac{n-1}{2}-J}}{2J}
\frac{\c_k\left(\vartheta\right)}{\s_k\left(\vartheta\right)^{2J}}\right)
,\\
\texttt{R}_{n-{\rm even}} \!&=&\! -\sum ^{\frac{n}{2}}_{J=1}
\left[\prod^{\frac{n}{2}-J}_{i=1}\frac{n-2i}{n-2i+1}\right]
\frac{k^{\frac{n}{2}-J}}{2J-1}\frac{\c_k\left(\vartheta\right)}{\s_k\left(\vartheta\right)^{2J-1}}
,\end{eqnarray*}
where we treat $k$ to the power of zero as $k^0=1$ for all admissible values of $k.$

We only consider $D\geq4$ and hence $n~=~\beta~=~(D-2)~\geq~2$ in the cases of interest.

\begin{figure}[p]
\centering
\includegraphics[scale=0.95]{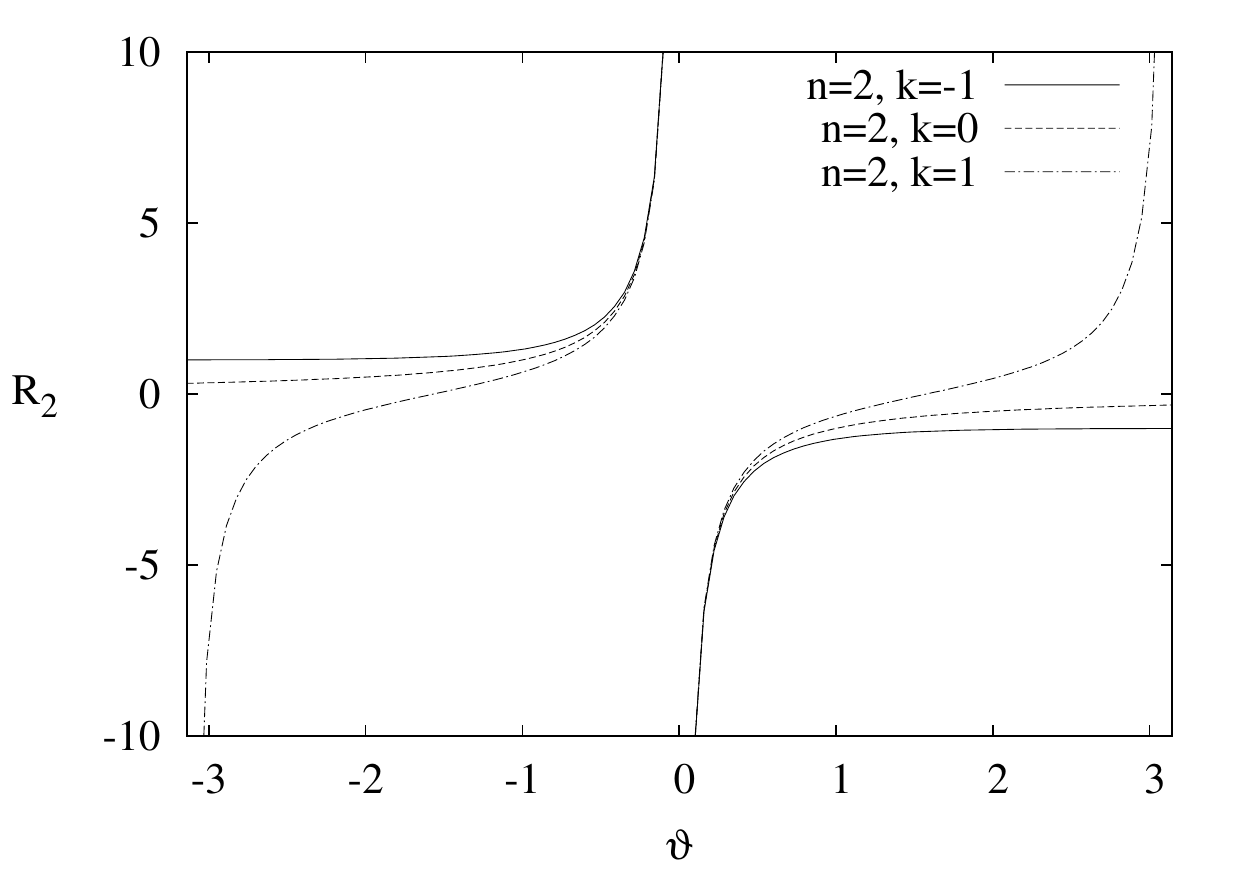}
\caption{Functions $\texttt{R}_n$ for $n=2$, $k=-1$, $k=0$, $k=1$.}
\label{gr1}
\end{figure}

\begin{figure}[p]
\begin{center}
\includegraphics[scale=0.95]{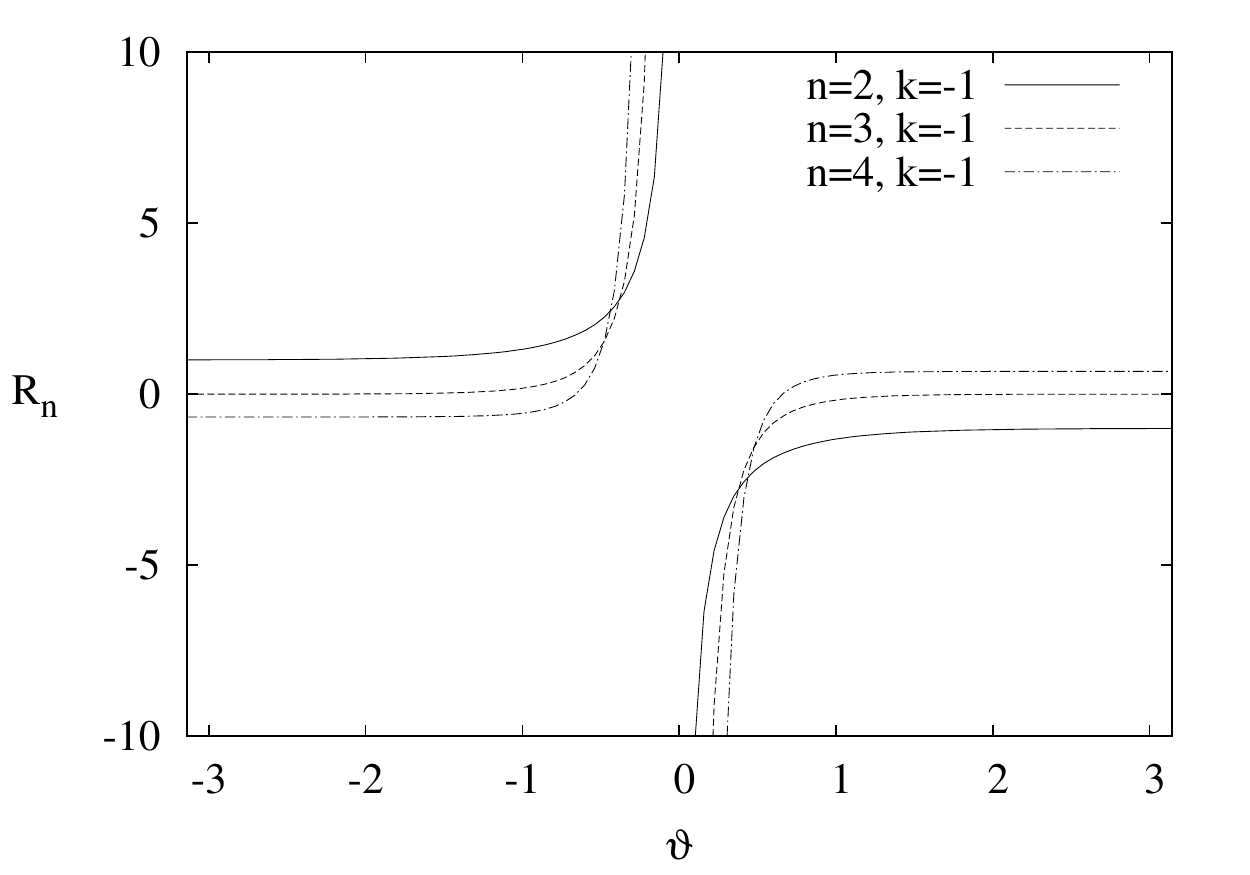}
 \end{center}
 \caption{Functions $\texttt{R}_n$ for $n=2$, $n=3$, $n=4$, $k=-1$. }
\label{gr2}
\end{figure}

In the case $k~=~0$ we can write the function $\texttt{R}_n$ in the form
$$\texttt{R}_{n\geq2}=
\sgn(\vartheta)^{n}\frac{1}{1-n}\frac{1}{\vartheta^{n-1}}.$$
The $n~=~2$ (i.e. $D~=~4$) total metric function $\texttt{G}$ takes the form
\begin{equation}
\texttt{G}(t,\vartheta) =
\texttt{T}(t) + C_1\frac{\c_k(\vartheta)}{\s_k(\vartheta)},\;
\texttt{T}(t) = Hct + C_2.
\label{G.D.4}\end{equation}
A few representatives of the $\texttt{R}_n$-functions are plotted in Figures~\ref{gr1} and~\ref{gr2}.

These figures indicate that the higher dimensional ($n~>~2$) solutions are very similar to the $D=4$ ($n~=~2$) example as can be seen from the behaviour of $\texttt{R}_n$ in Figure~\ref{gr2}.


Asymptotic behaviour of $\texttt{R}_n$ can be determined using
\begin{eqnarray*}
\lim_{\vartheta\rightarrow \pi^{-}}\frac{\s_1}{\pi-\vartheta} =
-\lim_{\vartheta\rightarrow \pi^{-}}\c_1 =
\lim_{\vartheta\rightarrow 0^{+}}\frac{\s_k}{\vartheta} =
\lim_{\vartheta\rightarrow 0^{+}}\c_k =
\lim_{\vartheta\rightarrow\infty}\frac{\s_{-1}}{\c_{-1}} = 1,
\end{eqnarray*}
\begin{eqnarray*}
\lim_{\vartheta\rightarrow\infty}\s_0 =
\lim_{\vartheta\rightarrow\infty}\s_{-1} =
\lim_{\vartheta\rightarrow\infty}\c_{-1} = \infty,
\end{eqnarray*}
argument $\vartheta$ is understood in all $\s_k$ and $\c_k$ in the above limits.
The limits $\lim\limits_{\vartheta\rightarrow\infty}s_1$ and $\lim\limits_{\vartheta\rightarrow\infty}c_1$ are not defined.

The above mentioned limits help us to calculate corresponding limits of the functions $\texttt{R}_n.$
The case of $\vartheta~\rightarrow~0^+$ is simple to treat and gives
$$k\in\{0,\pm 1\}, n\geq 2:
\left|\texttt{R}_n(\vartheta\rightarrow 0^+)\right|\approx
\frac{1}{(n-1)\vartheta^{n-1}} 
\rightarrow\infty.$$


Examination of the remaining case of $\vartheta~\rightarrow~\infty$ splits into several sub--cases depending on the value of both $k$ and $n.$

We note that $k~=~1$ cannot be treated since the corresponding limits of $\s_1$ and $\c_1$ do not exist.
\begin{eqnarray*}
k = 0: & \texttt{R}_n(\vartheta\rightarrow\infty) &
=
\def\arraystretch{0.6}\begin{array}{rl}
-\frac{1}{n-1}\frac{1}{\vartheta^{n-1}}\rightarrow 0 & n\geq 2
\end{array}
,\\
k = -1: & \texttt{R}_{n}(\vartheta\rightarrow\infty) &
= \left\{\def\arraystretch{1.2}\begin{array}{ll}
0 & n-{\rm odd}
\\
\prod\limits^{\frac{n}{2}-1}_{i=1}\left(\frac{n-2i}{n-2i+1}\right)\left(-1\right)^{\frac{n}{2}}& n-{\rm even}
\end{array}\right.
.\end{eqnarray*}
In the case of $k=1,$ it is useful to examine the limit $\vartheta~\rightarrow~\pi^{-}$ of the function $\texttt{R}_{n}$
\begin{eqnarray*}
k = 1: & \left|\texttt{R}_n(\vartheta\rightarrow \pi^{-})\right| &
\sim\left\{\def\arraystretch{0.6}\begin{array}{rl}
\frac{1}{\left(\pi-\vartheta\right)^{n-1}} & n\geq 2
\end{array}\right\}\rightarrow\infty
.\end{eqnarray*}

Examination of the function $\texttt{R}_{n}$ reveals that in the case of $k~=~1,$ the function satisfies $\texttt{R}_{n}(\vartheta=\frac{\pi}{2})=0.$

\bigskip
The metric function $\texttt{G}$ is singular at $\vartheta=0,$ so is $g_{11}~=~-\texttt{G}^{2\xi_D},$ irrespective of the value of the parameter $k$ and also at $\vartheta=\pi$ only for $k=1.$

In the case of $k=1,$ the function $\texttt{G}$ also diverges in $\vartheta~\rightarrow~\pi^{-},$ but this time it approaches minus infinity.
This can be proved using the simple relation between the limits $\vartheta\rightarrow~0^+$ and $\vartheta\rightarrow~\pi^-$ following from
$$\sin(\vartheta) = \sin(\pi-\vartheta),\; \cos(\vartheta) = -\cos(\pi-\vartheta)$$
and the formulae for the function $\texttt{R}_n.$
It turns out that these two limits differ in sign change only.

\medskip
Further analysis of some Riemann tensor related scalar invariants, presented in the next section, reveals singularities too, though at different values of $\vartheta.$

%
%

\medskip
Using the Misner--Sharp mass~\cite{Csizmadia:2009dm}
$m_{\rm MS}~=~\displaystyle\frac{r_{\rm MS}}{2}\left\{1+\nabla^{\mu}r_{\rm MS}\nabla_{\mu}r_{\rm MS}\right\}$ to our solution yields
\begin{equation}
m_{\rm MS} = \frac{r_{\rm MS}}{2}\left\{1+(\xi_D\dot{\texttt{T}})^2r^2_{\rm MS}-\left[
\c_k-\frac{\xi_DC_1\sgn^{D-2}(\s_k)}{r^{D-3}_{\rm MS}}
\right]^2\right\}
,\label{Misner.Sharp.Mass}\end{equation}
with $r_{\rm MS}^{2}$ being the "metric coefficient" at $\d\Omega^2,$ i.e. $r_{\rm MS}~=~\texttt{G}^{\xi_D}\s_k.$
The above results~\eqref{Misner.Sharp.Mass}, as well as the subsequent analysis, holds for the arbitrary function $\texttt{T}(t).$

The Misner--Sharp mass~\eqref{Misner.Sharp.Mass} is a rather complicated function of both $t$ and $\vartheta$ but we present some limits in Table~\ref{Misner.Sharp.Mass.Limits}.
The limits include $\vartheta_0$ which is a root of $\texttt{G}~=~0.$

\smallskip
For comparison, the Misner-Sharp mass calculated for Reissner--Nordstr\"{o}m~\cite{Stephani2003}
gives $m_{\rm MS}~=~\left(R_s~-~\frac{R^2_Q}{r}\right)/2.$

All $k~=~\dot{\texttt{T}}~=~0,$ $\texttt{T}~=~1$ limits, with $D~=~4,$ in the Table~\ref{Misner.Sharp.Mass.Limits} correspond to extremely charged Reissner--Nordstr\"{o}m (with $R_s=2R_Q$) so that $C_1$ can be identified with $R_Q$ of that metric.
This suggests that $C_1$ corresponds to the mass and charge of our solution as well.

\begin{table}[h]\begin{center}\def\arraystretch{1.9}\begin{tabular}{|l|l|l|l|l|l|}\hline
${\rm limits}$ & $k$ & $\texttt{T}(t)$ & $\dot{\texttt{T}}(t)$ & $D\geq 4$ & $2m_{\rm MS}$
\\\hline\hline
$\vartheta\rightarrow 0^+$ & -- & -- & -- & -- & $(\xi_DC_1)^{\xi_D}\left[
1+(\xi_D\dot{\texttt{T}})^2(\xi_DC_1)^{2\xi_D}
\right]$
\\\hline
$\vartheta\rightarrow\infty$ & $0$ & $\neq 0$ & $=0$ & $D=4$ & $2C_1$
\\\cline{5-6}
& & & & $D>4$ & $0$
\\\cline{4-6}
& & & $\neq 0$ & -- & $\infty$
\\\cline{3-6}
& & $= 0$ & $\neq 0$ & -- & $(\xi_DC_1)^{\xi_D}\left[
1+(\xi_D\dot{\texttt{T}})^2(\xi_DC_1)^{2\xi_D}
\right]$
\\\hline
$\vartheta\rightarrow\vartheta_0$ & -- & -- & -- & -- & $-\infty$
\\\hline
\end{tabular}\end{center}
\caption{
Various selected limits of $m_{\rm MS}$ as a function of $\vartheta.$
We assume the root $\vartheta_0$ of $\texttt{G}~=~0$ satisfies $\vartheta_0\not\in\{0, \pm\infty\}.$}
\label{Misner.Sharp.Mass.Limits}\end{table}






\section{Singularities and horizons}
It is important to distinguish between different types of singularities of various geometrical objects.

Analysis of singular behaviour independent of coordinate system choice uses scalar quantities.
We call singularities of scalars real singularities.

In addition to the metric components already examined, we will investigate three curvature scalars -- the Ricci scalar $R=R_{\alpha\beta}g^{\alpha\beta}$, Ricci square $R_{c}^2=R_{\alpha\beta}R^{\alpha\beta}$ and Kretschmann scalar \cite{Lemos:1994xp} (Riemann square) $R_{m}^2=R_{\alpha\beta\gamma\delta}R^{\alpha\beta\gamma\delta}.$

\medskip

\medskip
The three curvature scalars 
are given by
\begin{equation}
R=\frac{\xi_{D}}{\texttt{G}^{\alpha_{D1}}}\left(2A+\frac{(D-4)\texttt{G}_{,\vartheta}^2}{\texttt{G}}\right)-k\frac{\beta_{D1}}{\texttt{G}^{2\xi_{D}}}
-\alpha_{D1}\left(2\ddot{\texttt{G}}\texttt{G}+\alpha_{D0}\dot{\texttt{G}}^2\right),\label{19z}
\end{equation}

\begin{equation}
\def\arraystretch{1.5}\begin{array}{rcl}
R_{c}^2&=&\left[\frac{\texttt{G}A-\texttt{G}_{,\vartheta}^2}{\texttt{G}^{2\alpha_{D2}}}+\alpha_{D1}(\dot{\texttt{G}}\texttt{G}+\xi_{D}\dot{\texttt{G}}^2)\right]^2\\
&+&\left[\frac{\texttt{G}A+\texttt{G}_{,\vartheta}^2}{\texttt{G}^{2\alpha_{D2}}}-\left(k\frac{D-2}{\texttt{G}^{2\xi_{D}}}+B\right)\right]^2\\
&+&(D-2)\left[\frac{\texttt{G}A-\texttt{G}_{,\vartheta}^2}{\texttt{G}^{2\alpha_{D2}}}-\left(k\frac{D-2}{\texttt{G}^{2\xi_{D}}}+B\right)\right]^2
,\end{array}\label{20z}
\end{equation}

\begin{equation}
\def\arraystretch{1.5}\begin{array}{rcl}
R_{m}^2&=&\frac{4}{\texttt{G}^{4\alpha_{D2}}}\left\lbrace \left[\texttt{G}_{,\vartheta\vartheta}\texttt{G}-(2D-5)\xi_{D} \texttt{G}_{,\vartheta}^2+C\texttt{G}^{2\alpha_{D2}}\right]^2\right.
\\
&+&(D-2)\left[\texttt{G}_{,\vartheta}\frac{\c_k(\vartheta)}{\s_k(\vartheta)}\texttt{G}+\texttt{G}_{,\vartheta}^2\xi_{D}+C\texttt{G}^{2\alpha_{D2}}\right]^2
\\
&+&(D-2)\left[\left(\texttt{G}_{,\vartheta\vartheta}+\texttt{G}_{,\vartheta}\frac{\c_k(\vartheta)}{\s_k(\vartheta)}-\frac{\texttt{G}_{,\vartheta}^2}{\texttt{G}}\right)\texttt{G}\xi_{D}-k\texttt{G}^2-\dot{\texttt{G}}^2\xi_{D}^2\texttt{G}^{2\alpha_{D2}}\right]^2
\\
&+&\left. \frac{\beta_{D2}}{2}\left[\left(2\texttt{G}_{,\vartheta}\frac{\c_k(\vartheta)}{\s_k(\vartheta)}+\frac{\texttt{G}_{,\vartheta}^2\xi_{D}}{\texttt{G}}\right)\texttt{G}\xi_{D}-k\texttt{G}^2-\dot{\texttt{G}}^2\xi_{D}^2\texttt{G}^{2\alpha_{D2}}\right]^2\right\rbrace
,\end{array}\label{21z}
\end{equation}
where we have used some abbreviations for the sake of brevity
$$A=\left(\Delta_{(\gamma)}\texttt{G}\right)=\texttt{G}_{,\vartheta\vartheta}
+(D-2)\texttt{G}_{,\vartheta}\frac{\c_k(\vartheta)}{\s_k(\vartheta)},\;
B=\xi_{D}[\ddot{\texttt{G}}\texttt{G}+\alpha_{D1}\dot{\texttt{G}}^2],\;
C=\xi_{D}[\ddot{\texttt{G}}\texttt{G}+\xi_{D}\dot{\texttt{G}}^2].$$
The abbreviations can be simplified in the case of $\texttt{T}=Hct$ and source--free region solution~\eqref{18z}
$$A=0,\; B=\alpha_{D1}\xi_{D} H^2,\; C=\xi_{D}^2H^2.$$

Let us examine the scalars at $\vartheta~\rightarrow~0^{+}$ at a specific time $t=t_0.$
Functions $\texttt{T}=Hct_0,$ $\dot{\texttt{T}}=H,$ $\ddot{\texttt{T}}=0$ represent only constants in the limit of interest.
The limit $\vartheta~\rightarrow~0$ implies
$$\vartheta\rightarrow 0:\;
\s_k\rightarrow\vartheta,\;\c_k\rightarrow1,\; n\geq 2:
\texttt{R}_{n}\rightarrow\frac{\sgn(\vartheta)^n}{(1-n)\vartheta^{n-1}},$$
which in turn yields
$$
\texttt{R}\approx \texttt{G}\rightarrow\frac{\sgn(\vartheta)^{D-2}\xi_{D} C_1}{\vartheta^{D-3}},\;
\texttt{G}_{,\vartheta}\rightarrow
-\frac{C_1\sgn(\vartheta)^{D-2}}{\vartheta^{D-2}},\;
\texttt{G}_{,\vartheta\vartheta}\rightarrow\left(D-2\right)\frac{C_1\sgn(\vartheta)^{D-2}}{\vartheta^{D-1}}$$
in the case of $D~\geq~4.$
We will use the limit $\displaystyle\frac{\texttt{G}_{,\vartheta}}{\texttt{G}^{\alpha_{D2}}}\rightarrow-C_1\left(\frac{D-3}{C_1\sgn(\vartheta)}\right)^{\alpha_{D2}}.$

Limits $\vartheta~\rightarrow~0^{+}$ of the three curvature scalars can be written in the form
\begin{equation}
\begin{array}{rcl}
R&=&\alpha_{D4}C_1^2(\frac{D-3}{C_1})^{2\alpha_{D2}}-\alpha_{D1}\alpha_{D0}H^2,
\\
R_{c}^2&=&2\left[H^2\alpha_{D1}\xi_{D}-C_1^2(\frac{D-3}{C_1})^{2\alpha_{D2}}\right]^2+(D-2)\left[H^2\alpha_{D1}\xi_{D}
+C_1^2(\frac{D-3}{C_1})^{2\alpha_{D2}}\right]^2,
\\
R_{m}^2&=&4\left\lbrace \left[(D-3)C_1\left(\frac{D-3}{C_1}\right)^{\alpha_{D1}}-H^2\xi_{D}^2\right]^2+2(D-2)\left[H\xi_{D}\right]^4\right.
\\
&+&\left. \frac{\beta_{D2}}{2}
\left[C_1\xi_{D}\left(\frac{D-3}{C_1}\right)^{\alpha_{D1}}+H^2\xi_{D}^2\right]^2\right\rbrace .
\end{array}\label{22z}
\end{equation}
All the three scalars converge at the limit $\vartheta\rightarrow0^+$ into constants dependent on $C_1,$ the Hubble constant and the dimension of space-time.
This indicates that there is no physical singularity at this point.

The above outlined relation between the limits $\vartheta~\rightarrow~\pi^{-}$ and $\vartheta~\rightarrow~0^+$ in the case of $k=1$ implies the curvature scalars are not singular at $\vartheta~=~\pi.$ 

\medskip
Another important limit to explore is $\texttt{G}\rightarrow0^{+}.$
We denote the root(s) of $0~=~\texttt{G}(\vartheta)$ by $\vartheta_0.$

In order to have the Ricci scalar non-divergent one must impose
\begin{equation}
\left(\frac{\texttt{G}_{,\vartheta}}{\texttt{G}}\right)^2
\stackrel{\vartheta\rightarrow\vartheta_0
}{\longrightarrow}
\frac{k\beta_{D1}}{\alpha_{D4}}\Rightarrow
\pm\sqrt{\frac{k\beta_{D1}}{\alpha_{D4}}} =
\left.\frac{\texttt{G}_{,\vartheta}}{\texttt{G}}\right|_{\vartheta_0} =
-\left.\frac{C_1}{|\s^{D-2}_k|\texttt{G}}\right|_{\vartheta_0}
.\label{Ricci.scalar.limit.G.0}\end{equation}

One has to have $C_1~=~0$ or $|\s_k|\rightarrow\infty$ to achieve a finite ratio of $\texttt{G}_{,\vartheta}/\texttt{G}$ at $\vartheta_0.$
The first possibility is omitted since it leads to a rather trivial metric function $\texttt{G}.$
The second one, $|\s_k|\rightarrow\infty,$ can only be achieved if both $k\in\{0,-1\}$ and $1/\vartheta_0~=~0$ hold true.


If $k~=~-1,$ then the first relation in equation~\eqref{Ricci.scalar.limit.G.0} yields
$$0\leq\left.\left(\frac{\texttt{G}_{,\vartheta}}{\texttt{G}}\right)^2
\right|_{\vartheta_0} = -\frac{\beta_{D1}}{\alpha_{D4}}<0,$$
i.e. the condition expressed in~\eqref{Ricci.scalar.limit.G.0} cannot be satisfied.

The relation between $\texttt{G}$ and its $\vartheta-$derivative is satisfied in the case of both $k~=~0$ and $1/\vartheta_0~=~0.$

\smallskip
Thus it follows $k~\neq~0$ or both $k~=~0$ and $1/\vartheta_0~\neq~0$ implying that the Ricci scalar diverges for $\texttt{G}\rightarrow~0^+.$

\medskip
Consider $C_2=0,$ $C_1>0$ and a positive root $H$ from equation~\eqref{Kastor.Traschen.like} and $t=t_0>0;$ then $\texttt{T}_0=Hct_0$ is a positive constant.

We would like to examine whether the limit $\texttt{G}\rightarrow0^{+}$ can be achieved for an arbitrary positive value of $t_0.$
$\texttt{G}~=~0$ for some $\vartheta_0$ means $\texttt{R}_{D-2}(\vartheta_0)=-\frac{\texttt{R}(\vartheta_0)}{C_1}=\frac{\texttt{T}_0}{C_1}.$
This implies that if $\texttt{T}_0$ is positive, then $\texttt{R}_{D-2}(\vartheta_0)$ must be also positive for a given assumption on the constant $C_1.$

Figure~\ref{gr1} makes clear that in the case of $k=0,$ the function $\texttt{R}_{D-2}(\vartheta_0)$ is positive if $\vartheta_0$ is negative and conversely.
The above holds true for all considered values of $k$ for a sufficiently small $\vartheta_0$ (not necessarily infinitesimal).

\medskip

One can calculate the spatial volume bounded by surfaces $\vartheta~=~\vartheta_0$ and $\vartheta~=~0$
$$V_{[\vartheta_0,0]} = 
\Omega\int^0_{\vartheta_0}\left|\texttt{G}\right|^{\alpha_{D1}}\left|\s_k\right|^{D-2}\d\vartheta$$
is infinite due to the singular behaviour at the $\vartheta~=~0$ surface.
The constant $\Omega$ denotes integral over the angular variables.

The same argument implies that all spatial volumes including $\vartheta~=~0$ in the integration domain over $\vartheta$ are infinite.

Let us consider the case of $D~=~4$, $\texttt{T}>0$, $\s_k>0$.
We can calculate the time change of (total) spatial volume of the time slices as
\begin{eqnarray*}
\mathcal{\dot{V}}_{D=4} &=&
3H\Omega\int\left(\texttt{T}\s_k+C_1\c_k\right)^{2}\d\vartheta =\\
&=&3H\Omega\left[C_1^{2}\vartheta+C_1\texttt{T}\s_k^{2}+\left(\texttt{T}^{2}-kC_1^{2}\right)\frac{1}{2k}\left(\vartheta-\s_k\c_k\right) \right].
\end{eqnarray*}
The above calculation involves a less trivial integral
$$\int\s^2_k\d\vartheta = \frac{1}{2k}\left[\vartheta - \s_k\c_k\right]\stackrel{"k\rightarrow 0"}{\longrightarrow}\frac{1}{3}\vartheta^3.$$
The above artificial limit of $k\rightarrow 0,$ treating $k$ as continuous, shows the result is valid for all $k\in\{-1,0,1\}.$

In the case of $k~=~1,$ $\vartheta\in[0,\pi],$ the result is finite and equal to
$$\mathcal{\dot{V}}_{D=4} = \frac{3\pi}{2}H\Omega\left[C_1^{2}+\texttt{T}^{2}\right].$$

We can interpret the above relation as follows -- the change of size of the universe depends on the "scale function" $\texttt{T},$ as in the FRWL case, and the constant $C_1$ also enters the formula.

The above conclusions can be generalised into higher dimensions.

\medskip
In general, there are values of $\texttt{T}_0$ for which one cannot achieve $\texttt{G}=0.$
These values are $\texttt{T}_0=0$ in the two cases of $k=0$ and both $k=-1$ and $D-$odd.
$$\frac{\texttt{T}_0}{C_1}\in\left[a,-a\right],\;
a=(-1)^{\frac{D}{2}+1}\prod^{\frac{D}{2}-2}_{i}\frac{D-2-2i}{D-1-2i}
$$
in the case of both $k=-1$ and $D~=~4m,$ $m\in\mathbb{N}.$

This follows from the limits of $\texttt{R}_n$ in Section~\ref{Extreme.k} and the fact that the function $\texttt{R}_n$ is monotonous on the intervals $(0,\infty)$ and $(-\infty,0).$
The latter is a direct consequence of the $\texttt{R}_n$ definition via $\texttt{R}$ in equation~\eqref{18z} so that in the case of $k~=~-1$ one has $\texttt{R}_{,\vartheta}\propto~1/\left|\sinh(\vartheta)\right|^{D-2}.$

The KT-FRWL solution bears more resemblance to an extremely charged Reissner--Nordstr\"{o}m solution (which is a concrete example of the KT-FRWL family of solutions), set $R_s~=~2R_Q$ in
\begin{equation}
\d s^2 = f(r)\d t^2 - f^{-1}(r)\d r^2 - r^2\d\Omega^2_2,\;
f(r) = 1-\frac{R_s}{r}+\frac{R^2_Q}{r^2}
,\label{Reissner.Nordstroem}\end{equation}
rather than to the Schwarzschild solution (obtained from the above formula by setting $R_Q~=~0$).

$$\d s^2 =\left(1-\frac{R_{Q}}{r}\right)^2\left(c\d t\right)^2
-\frac{1}{\left(1-\frac{R_{Q}}{r}\right)^2}\d r^2-r^2\d\Omega^2_2
,$$
where $R_{Q}$ is a constant dependent on the mass and charge of the black hole.

The above form of the extremely charged Reissner--Nordstr\"{o}m solution can be obtained from the following simple solution contained within the KT-FRWL ansatz~\eqref{MP.FRWL.metric}, see also~\eqref{metric.k.cases}, with $\texttt{G}=1+\frac{R_{Q}}{\vartheta}$
$$\d s^2 =\frac{1}{\left(1+\frac{R_{Q}}{\vartheta}\right)^2}\left(c\d t\right)^2
-\left(1+\frac{R_{Q}}{\vartheta}\right)^2\left(\d\vartheta^2+\vartheta^2\d\Omega^2_2
\right)$$
by transformation $r=R_{Q}+\vartheta.$

The Reissner--Nordstr\"{o}m singularities occur at $\vartheta=0$ (the coordinate singularity) and $\vartheta=-R_{Q}$ (the curvature singularity at $\texttt{G}~=~0$).

The Reissner--Nordstr\"{o}m domain $r\in(0,\infty)$ translates to $\vartheta\in(-R_Q,\infty).$ 
Thus it is meaningful to consider that the coordinate $\vartheta$ may be negative in our solution.
I.e. one should include the sub-interval $(\vartheta_0,0]$ into the domain of definition of the metric for one of the $\vartheta_0~<~0,$ if it exists, at which $\texttt{G}(\vartheta_0)~=~0$ holds.

\medskip
Definitions of black holes (BH) or white holes (WH) usually operate with the full knowledge of the causal structure of the space--time which is difficult to obtain.
The need for some local and more accessible definitions has lead to the introduction of so called trapped surfaces and horizons~\cite{Schinkel:2013zm, Chu:2010yu, Husain:2004yy, Krishnan:2013saa, Booth:2005qc, Schnetter:2005ea, Faraoni:2013aba}.
These surfaces are defined using congruences of two null vectors $k_+$ (outgoing) and $k_-$ (ingoing) emanating from the surface to which they are perpendicular.
Such congruences may correspond to the EM signal propagating from the said surface.
The definitions involve expansions $\theta_{\pm}$~\cite{Krishnan:2013saa, Booth:2005qc}
\begin{equation}
\theta_{\pm}=q^{\alpha\beta}\nabla_{\alpha}k_{\pm\beta},\;\;
q^{\alpha\beta}=g^{\alpha\beta}-k^{\alpha}_{+}k^{\beta}_{-}-k^{\alpha}_{-}k^{\beta}_{+}
.\label{ex}
\end{equation}

\begin{itemize}
\item
A marginally trapped surface (MTS) is a surface where 
$\theta_{+}\theta_{-}=0$ holds~\cite{Schnetter:2005ea}.

\item
A future trapping horizon (FTH) is the closure of a (usually a $D-1$--)surface foliated by MTS such on its $D-2$ "time slicings" $\theta_{-}<0,$ $\theta_{+}=0$
~\cite{Faraoni:2013aba}.

\item
The definition of a past
trapping horizon (PTH) is obtained by exchanging $k^{\alpha}_{+}$ a with $k^{\alpha}_{-}$ and reversing the signs of the inequalities, $\theta_{+}>0,$ $\theta_{-}=0.$
\end{itemize}

These two trapping horizons may provide definitions of black holes (FTH) or white holes (PTH) and they also include cosmological horizons as well.

\medskip
In our case of the solution~\eqref{MP.FRWL.metric} with a spherical symmetry, one has the $k_{\pm}$ given by~\eqref{WAND.maximal.beta} so that
$$q_{\alpha\beta}\d x^{\alpha}\d x^{\beta}=-\s_{k}^{2}\texttt{G}^{2\xi_{D}}\d \Omega^{2}.$$
Expansions for metric \eqref{MP.FRWL.metric} are given by $\theta_{\pm}=\frac{1}{2}k^{\beta}_{\pm}\sum_{\alpha}q^{\alpha\alpha}q_{\alpha\alpha,\beta}$
\begin{eqnarray*}
\theta_{\pm} 
&=&\frac{\alpha_{D2}}{\sqrt{2}}\left[\dot{\texttt{T}}
\pm\texttt{G}^{-\xi_{D}}\left(\frac{\texttt{R}_{,\vartheta}}{\texttt{G}}+\left(D-3\right)\frac{\c_{k}}{\s_{k}}\right)\right]
.\end{eqnarray*}

$\theta_{\pm}=0$ implies both $\theta_{\mp}=\alpha_{D2}\sqrt{2}\dot{\texttt{T}}$ and $\dot{\texttt{T}}=\mp\texttt{G}^{-\xi_{D}}\left(\frac{\texttt{R}_{,\vartheta}}{\texttt{G}}+\left(D-3\right)\frac{\c_{k}}{\s_{k}}\right)$ must hold.
The latter relation should be solved with respect to $\vartheta$ to obtain time-dependent positions of the horizon(s).
A double null surface ($\theta_+=\theta_-=0$) can only be present if $\dot{\texttt{T}}~=~0,$ e.g. the horizon of extremal Reissner--Nordstr\"{o}m.


Kastor and Traschen have found~\cite{Kastor:1992nn} that the $D~=~4$, $\ddot{\texttt{T}}~=~k~=~0$ singularity is a white hole if $\dot{\texttt{T}}>0$ and it is a black hole if $\dot{\texttt{T}}<0.$

In a general case, we have found that $\dot{\texttt{T}}~>~0$ implies the presence of a PTH and $\dot{\texttt{T}}~<~0,$ on the other hand, implies the presence of an FTH.

Notice that the above analysis of horizons applies to the metric solution~\eqref{14} with an arbitrary time dependent function $\texttt{T}.$

\section{Energy conditions}\label{energy.conditions}
Let us consider the source~\eqref{2} of co--moving matter with $F_{0\vartheta}=-F_{\vartheta0}$ being the only nonzero components of the EM strength tensor.
The corresponding solution~\eqref{18z} has been found in the region where $\rho_{\rm e}~=~0,$ hence the energy conditions will be examined in the same region.


\medskip
Let us examine non--negativity of the scalar
\begin{equation}T_W\equiv p_{\alpha}w^{\alpha},\;
p_{\alpha}\equiv T_{\alpha\beta}w^{\beta},\;
w^{\beta}w_{\beta} = W\in\{0, 1\}
,\label{weak.null.in.one}\end{equation}
where $w^{\beta}$ is a normalised timelike future directed or null vector whose square equals $W.$
The choice of matter source implies
\begin{eqnarray*}
T_W &=& \left[\rho_{\rm m}c^2 + p\right]
\left(u_{\mu}w^{\mu}\right)^2 - pW + \varepsilon_0c^2\left[
\frac{1}{4}F_{\alpha\beta}F^{\alpha\beta}W
- F_{\mu\alpha}F^{\;\alpha}_{\nu}w^{\mu}w^{\nu}\right]
\\
&=& \left[\rho_{\rm m}c^2 + p\right]
g_{00}(w^0)^2
- pW + \varepsilon_0c^2
\left(F_{0\vartheta}\right)^2\texttt{G}^2\left[
\frac{W}{2\texttt{G}^{2\xi_D}} + \sum_{l\geq 2}\gamma_{ll}
\left(w^l\right)^2\right]
,\end{eqnarray*}
where we have assumed that the metric $\gamma_{ij}$ is diagonal.

The last term in the above relation is manifestly non--negative.

Restricting to the $\ddot{\texttt{G}}~=~0$ solution, characterised by $\lambda~=~0$ version of ~\eqref{Kastor.Traschen.like}, in the region of $\rho_{\rm e}~=~0,$ one obtains
$$T_W = \frac{k}{K}\frac{\left(D-2\right)}{\texttt{G}^{2\xi_D}}\left[
\frac{(w^0)^2}{\texttt{G}^{2}} + \frac{D-3}{2}W\right]
+ \varepsilon_0c^2
\left(F_{0\vartheta}\right)^2\texttt{G}^2\left[
\frac{W}{2\texttt{G}^{2\xi_D}} + \sum_{l\geq 2}\gamma_{ll}
\left(w^l\right)^2\right]
.$$
We see that if $k~\in\{0,1\},$ then $T_W~\geq~0$~for arbitrary $W$ (which is assumed to be non--negative by definition~\eqref{weak.null.in.one}).

If $k~=~-1,$ then select vector $w^{\mu}$ so that the only non--zero components are $w^0$ and $w^1.$
These two components can be made arbitrarily large without affecting the normalisation of the vector $w^{\mu}.$
With $w^0$ being large enough, the first term in the last relation for $T_W$ is negative enough so that the $T_W~<~0.$

\smallskip
The weak energy condition can be formulated using the scalar $T_W$ as $T_1~\geq~0$ and the null energy condition is given by $T_0~\geq~0.$
We see that these two conditions are violated in the case of $k~=~-1.$



\medskip
The fact that both $g_{\mu\nu}$ and $T_{\mu\nu}$ are diagonal simplifies the dominant energy condition $p^{\mu}p_{\mu}~\geq~0$ to
\begin{equation}p^{\mu}p_{\mu} = W\left(T^0_0\right)^2+\texttt{G}^{2\xi_D}\sum^{D-1}_{i=1}\gamma_{ii}\left(w^i\right)^2\left[
\left(T^0_0\right)^2-\left(T^i_i\right)^2\right]\geq 0
.\label{dominant}\end{equation}
The condition is identically satisfied if the square bracket term -- that is actually independent of $i$ -- is non--negative.
If it is negative, then one can proceed analogously to the examination of $T_W~\geq~0$ in the case of $k~=~-1.$
With $w^1$ being large enough, the term with the sum in the relation~\eqref{dominant} is negative enough so that the dominant energy condition is violated.

Thus the sign of the square bracket term in the relation~\eqref{dominant} determines whether the condition is violated or not, i.e. the condition holds if the following inequality holds
$$\left|T^0_0\right|\geq \left|T^i_i\right|;\;
T^0_0 = \rho_{\rm m}c^2+\frac{1}{2}\varepsilon_0c^2\left(F_{0\vartheta}\right)^2\texttt{G}^{2-2\xi_D},\;
T^i_i = -p+\frac{1}{2}\varepsilon_0c^2\left(F_{0\vartheta}\right)^2\texttt{G}^{2-2\xi_D}.$$
Using the $\lambda~=~0$ version of formula~\eqref{Kastor.Traschen.like}, one can find that the condition is identically satisfied if $k\in\{0,~1\}$ and that in the case of $k~=~-1$ the pressure is bounded from below $p\geq\frac{1}{\alpha_{D1}+1}\varepsilon_0c^2\left(F_{0\vartheta}\right)^2\texttt{G}^{2-2\xi_D}.$

The vector $p^{\mu}$ is future directed if $p^0~=~T^0_0w^0~>~0$ holds.
This is obviously the case of both $k\in\{0,1\}$ and $w^0~>~0.$
In the case of $k~=~-1,$ the $p^{\mu}$ is not necessarily future directed.

%
%

\medskip
The $w^{\mu}~=~u^{\mu}$ strong energy condition can be written as
\begin{equation}
\begin{array}{rcl}
R_{\alpha\beta}u^{\alpha}u^{\beta}&=&KT_{\alpha\beta}u^{\alpha}u^{\beta}-\frac{1}{D-2}\left(KT^{\alpha}_{\;\;\alpha}+2\Lambda_{\rm cosm}\right)
\\
&=&K\frac{1}{\alpha_{D2}}\rho_{\rm m}c^2+K\frac{\alpha_{D1}}{\alpha_{D2}}p+\left(\frac{\texttt{G}_{,\vartheta}}{\texttt{G}^{\alpha_{D2}}}\right)^2-\frac{2}{D-2}\Lambda_{\rm cosm}
\\
&=&K\frac{1}{\alpha_{D2}}\rho_{\rm me}c^2-\frac{2}{D-2}\Lambda_{\rm cosm}+\left(\frac{\texttt{G}_{,\vartheta}}{\texttt{G}^{\alpha_{D2}}}\right)^2
\geq 0,
\end{array}
\label{24z}\end{equation}
where we have used
$$\rho_{\rm m}c^2+\alpha_{D1}p=\rho_{\rm me}c^2+\left(\rho_{\rm mp}c^2+\alpha_{D1}p\right)=\rho_{\rm me}c^2.$$
The above relation among the fluid source fields holds for the $\beta~=~(D~-~2)$ and $\lambda~=~0$ particular solution of~\eqref{Kastor.Traschen.like}.

The charge density of these particular solutions is nonzero only at curvature scalars singularities.
The strong energy condition outside the singularities ($\rho_{\rm me}~=~0$) is
$$\left(\frac{\texttt{G}_{,\vartheta}}{\texttt{G}^{\alpha_{D2}}}\right)^2-\frac{2}{D-2}\Lambda_{\rm cosm} =
\left(\frac{\texttt{G}_{,\vartheta}}{\texttt{G}^{\alpha_{D2}}}\right)^2-\frac{\alpha_{D1}}{D-3}H^2
\geq0,$$
where we have used an equation describing the relationship between the Hubble constant and the cosmological constant presented in~\eqref{Kastor.Traschen.like}.

The inequality can be written as
\begin{equation}
\left|\frac{\texttt{G}_{,\vartheta}}{\texttt{G}^{\alpha_{D2}}}\right|\geq\sqrt{\frac{\alpha_{D1}}{D-3}}\left|H\right|
.\label{strong.limit}\end{equation}
The considered solution need not necessarily satisfy this condition for every value of $t$ or $\vartheta.$

Let us consider an illustrative example of the strong energy condition being violated using concrete numerical values.

The value of the Hubble constant is approximately $\mathcal{H}\approx10^{-18}\rm{s}^{-1}.$
The related constant $H=\frac{\mathcal{H}}{c}$ is about $H\approx10^{-26}\rm{m}^{-1}.$
If we choose $D=4,$ $k=0$ and if we further assume the spatial-part in the metric function $\texttt{G}$ dominates, $Hct+C_2\ll\frac{C_1}{\vartheta},$ then the condition~\eqref{strong.limit} can be written as
$$\frac{1}{C_1}>10^{-26}\rm{m}^{-1}.$$
The above KT limit allows us to write the constant of integration $C_1,$ related to the Schwarzschild radius, in form $C_1\approx 1.5\cdot 10^3M/M_{\rm Sun}$~m, where $M_{\rm Sun}$ is mass of the Sun and $M$ is mass of the charged matter source considered.
The above reduced inequality makes obvious that violating the strong energy condition requires that $M$ is larger than the mass of the entire visible Universe $M_{\rm Univ}\approx 1.5\cdot10^{22}M_{\rm Sun}.$

\section{Conclusions}\label{few.final.words}
In this work, we have studied a cosmologically inspired KT type solution in a higher dimensional space--time with a two component co--moving fluid matter source -- charged dust (a KT motivated contribution) and perfect fluid (an added "cosmological" substratum).
The choice of matter source enabled us to proceed analogously to the case of an MP solution.

We have written down the Einstein--Maxwell equation corresponding to both our ansatz metric and selected matter source, thus relating the space--time geometry with the matter fields.
Then, we turned our attention to finding and studying a class of exact spherically symmetric solutions generalising Reissner--Nordstr\"{o}m black hole solution.
Examination of a rather general solution is supplemented by a more detailed analysis of the $\ddot{\texttt{G}}~=~0$ case.

The behaviour of space--time's geometry, especially at $\texttt{G}$ either $0$ or $\infty$ at which some of the metric coefficients behave oddly, has been examined using curvature scalars at most quadratic in Riemann tensor and its metric contractions.
It was found that the curvature scalar singularities are located at $\texttt{G}~=~0.$

Analysis of trapped surfaces and horizons followed showing that the sign of $\dot{\texttt{T}}$ determines the type of the trapped horizon, a result generalising the work~\cite{Kastor:1992nn} where the sign of the same quantity distinguished between black and white hole space--times.

We have also discussed the validity of the energy conditions proving that the dominant, weak and null energy conditions are violated in the case of $k~=~-1$ and hold in $k=\left\lbrace0,1\right\rbrace$.
The strong energy condition can also be violated in some regions of the space--time irrespective of $k.$

\smallskip
A proposed generalisation of our previous work~\cite{Cermak:2012pt} has also been addressed in Section~\ref{no.shells} by considering a charged dust shell matter source with the conclusion that such a solution with a spatial metric $\gamma_{ij}$ given by~\eqref{metric.k.and.flat} cannot exist unless $k~=~0.$

\medskip
It would be interesting to find a solution to the Einstein--Maxwell equations with the metric function $\texttt{G}$ having a less trivial time dependence.
We note that some of the analysis present in this work already applies in the general case as well.

\bigskip
{\bf Dedication:}
In loving memory of prof. RNDr. Jan Horsk\'{y}, DrSc. ($^{*}$1940 -- $\dagger$2012).

\bibliographystyle{unsrt}
\bibliography{citace}

\end{document}